\documentclass[fleqn,usenatbib]{mnras}

\usepackage{newtxtext,newtxmath}
\usepackage{multicol}
\usepackage{graphicx}

\usepackage[T1]{fontenc}
\usepackage{ae,aecompl}

\usepackage{soul}
\usepackage{color}
\DeclareRobustCommand{\hlcyan}[1]{{\sethlcolor{cyan}\hl{#1}}}
\renewcommand\hl[1]{#1} %
\usepackage{caption}
\usepackage{subcaption}
\usepackage[flushleft]{threeparttable}

\usepackage{graphicx}	
\usepackage{amsmath}	

\title[SMBHs in stripped nuclei]{\hl{The contribution of supermassive black holes in stripped nuclei to the supermassive black hole population of UCDs and galaxy clusters}}

\author[R. J. Mayes et al.]{Rebecca J. Mayes,$^{1}$\thanks{E-mail: r.mayes@uq.net.au}
Michael. J. Drinkwater,$^{1}$
Joel Pfeffer,$^{2}$ Holger Baumgardt,$^{1}$  
\\
$^{1}$School of Mathematics and Physics, University of Queensland, Brisbane, QLD 4072, Australia\\
$^{2}$International Centre for Radio Astronomy Research (ICRAR), M468, University of Western Australia, 35 Stirling Hwy, Crawley, WA 6009, Australia\\
}

\date{Accepted XXX. Received YYY; in original form ZZZ}

\pubyear{2021}

\begin{document}
\label{firstpage}
\pagerange{\pageref{firstpage}--\pageref{lastpage}}
\maketitle

\begin{abstract}

We use the hydrodynamic EAGLE simulation to predict the numbers and masses of supermassive black holes in \hl{remnant nuclei of disrupted galaxies (stripped nuclei)} and compare these to confirmed measurements of black holes in observed ultra-compact dwarf galaxies (UCDs). We find that black holes in stripped nuclei are consistent with the numbers and masses of those in observed UCDs. Approximately 50 per cent of stripped nuclei with M > 2~$\times$~10\textsuperscript{6}~\(\textup{M}_\odot\) should contain supermassive black holes. 
We further calculate \hl{how the presence of a black hole increases the dynamical mass of a stripped nucleus via the mass elevation ratio, $\Psi$ defined as the ratio of the kinematically derived mass to the expected mass from stellar population synthesis. We find} $\Psi$\textsubscript{sim} $= 1.51^{+0.06}_{-0.04}$ for M > 10\textsuperscript{7}~\(\textup{M}_\odot\) stripped nuclei, consistent with that of observed UCDs which have $\Psi$\textsubscript{obs} = $1.7 \pm 0.2$ above M > 10\textsuperscript{7}~\(\textup{M}_\odot\). We also find that the mass elevation ratios of stripped nuclei with supermassive black holes can explain the observed number of UCDs with elevated mass-to-light ratios. Finally, we predict the relative number of massive black holes in stripped nuclei and galaxy nuclei and find that stripped nuclei should increase the number of black holes in galaxy clusters by $30-100$ per cent, depending on the black hole occupation fraction of low-mass galaxies. We conclude that the population of supermassive black holes in UCDs represents a large and unaccounted-for portion of supermassive black holes in galaxy clusters.

\end{abstract}




\section{Introduction}

\label{section:intro}
Ultra-compact dwarf galaxies (UCDs) are intermediate in size and brightness between dwarf galaxies and globular clusters \citep{Mieske2008, Voggel2016}. UCDs were first discovered around the end of the 20th century in spectroscopic surveys of the Fornax cluster \citep{Hilker1999, Drinkwater2000} and have since been discovered in other clusters, \citep{Mieske_2009, Chiboucas2011, Zhang2015, liu2020generation}, in galaxy groups \citep{Evstig2007a, DaRocha2011, Madrid2013, De_B_rtoli_2020} and around relatively isolated galaxies  \citep{Hau200, Norris2011}. Generally, UCDs are more common in clusters than groups \citep{Evstigneeva_2007b} and the general field \citep{Liske2006}.

 
The method by which UCDs form is a matter of some debate, and there are a variety of competing formation theories. The two main scenarios are that they are either (1) high-mass globular clusters or (2) the nuclei of disrupted dwarf galaxies.

In the first scenario, they would be the high-mass end of the globular cluster mass-function around galaxies with rich GC systems \citep{Mieske2002, Mieske2012}. 

In the second scenario, they are the compact nuclei of dwarf galaxies that underwent galaxy threshing, leaving only the nucleus remaining \citep{Bassino1994, Bekki2001, Bekki2003, Drinkwater2003, Goerdt2008, Pfeffer2013, Pfeffer2014}. There is a growing body of evidence that both scenarios make up some portion of the UCD population, with bright UCDs above a mass of approximately 10\textsuperscript{7}~\(\textup{M}_\odot\) being predominantly stripped nuclei \citep{Pfeffer2014, Pfeffer2016, Mayes2020}, while UCDs with lower masses are likely massive globular clusters. 
UCDs have central velocity dispersions similar to dwarf galaxies, in the range of 20 < $\sigma$\textsubscript{0} < 50~km s\textsuperscript{-1}. They have dynamical masses of approximately 2~$\times$~10\textsuperscript{6} to 10\textsuperscript{8}~\(\textup{M}_\odot\)
 \citep{has2005, Hilker2008, Mieske2008, Mieske2013}.
 
The elevated dynamical mass-to-light ratios of UCDs have received significant attention in the literature. Approximately two-thirds of UCDs with M $>$ 10\textsuperscript{7}~\(\textup{M}_\odot\) and one-fifth of UCDs with masses between 2~$\times$~10\textsuperscript{6}~\(\textup{M}_\odot\) and 10\textsuperscript{7}~\(\textup{M}_\odot\) require additional dark mass to explain their dynamical mass-to-light ratios \citep{Mieske2013, Voggel_2019}. 

One explanation for these M/L ratios is that UCDs contain central supermassive black holes \hlcyan{(SMBHs)} in the mass range of 10\textsuperscript{5}~\(\textup{M}_\odot\) to 10\textsuperscript{7}~\(\textup{M}_\odot\), \hlcyan{making up} a few per cent of the UCD mass \citep{Mieske2013}.
A central black hole of a given mass will enhance the velocity dispersion of a UCD more than a uniform distribution of the same mass \citep{Merritt_2009}. It is well known that supermassive black holes are found in the centre of galaxies \citep{Kormendy1995, Magorrian1998, Ferrarese2000, Gebhardt2000, Marconi2003, haring2004} with black hole mass increasing with galaxy mass.

With stripped nuclei UCDs, the original galaxy masses are expected to be in the range of 10\textsuperscript{8}~\(\textup{M}_\odot\) to 10\textsuperscript{10}~\(\textup{M}_\odot\) \citep{Bekki2003, Pfeffer2013}. In this range, central black holes are expected to be less massive than their host nuclei and to \hlcyan{strongly affect} the velocity dispersion because they are located at the centre \citep{Graham_2013}. Thus, moderate-sized supermassive black holes could explain the observed elevated M/L ratios. 

\citet{Mieske2013} \hl{determined dynamical masses and M/L ratios of \mbox{$\approx$}50 extragalactic UCDs using} \hlcyan{ spectroscopic velocity dispersions and assuming old (13 Gyr) stellar populations} and found that central black holes with masses of around 10-15 per cent of the mass of the UCD could explain the elevated M/L ratios.\hlcyan{Black hole masses would range} from several 10\textsuperscript{5}~\(\textup{M}_\odot\) to several 10\textsuperscript{7}~\(\textup{M}_\odot\) and progenitor galaxy masses \hlcyan{would be} around 10\textsuperscript{9}~\(\textup{M}_\odot\). In this case, assuming a canonical \hl{\mbox{\citep[e.g.][]{Kroupa2001}}} stellar population, about half the UCD population may contain a black hole with sufficient mass to elevate the M/L ratio, with the majority of UCDs containing black holes being high-mass UCDs with M $>$ 10\textsuperscript{7}~\(\textup{M}_\odot\), which make up \hlcyan{\mbox{$\lesssim 10$}} per cent of the overall UCD population.

An important question in studying UCDs is how the numbers and masses of their central supermassive black holes compare to black holes within galaxy nuclei. \citet{Seth2014} and \citet{Voggel_2019} \hl{independently} estimated the contribution UCDs should make to the supermassive black hole density in the universe. \citet{Seth2014} \hl{examined the dynamical $M/L_V$ ratios of UCDs in the Fornax cluster and }concluded that \hlcyan{the inclusion of black holes contained in} UCDs could more than double the number of black holes in Fornax. \citet{Voggel_2019} \hl{compared the dynamical and stellar population masses of a compiled sample of UCDs from the literature and }estimated that UCDs should increase the total supermassive black hole number density by 8-32 per cent. Therefore UCDs could represent a substantial increase in the number density of supermassive black holes in the universe, thus increasing black hole tidal disruption events and mergers.


In \citet{Mayes2020} (hereafter Paper I), we presented the first model for UCD formation based on tidal stripping of dwarf galaxies in a cosmological hydrodynamical simulation\hlcyan{. We found }that numbers and radial distributions of simulated stripped nuclei in Virgo-sized galaxy clusters were consistent with the numbers and radial distributions of UCDs observed in the Virgo cluster. In \citet{Mayes2021} (hereafter Paper II), we further tested the model by using the baryonic component of the EAGLE simulation \citep{Crain_2015, Schaye2015} to determine the ages, colours and metallicities of our simulated stripped nuclei and found they were consistent with observed UCDs.

In this paper, we use the hydrodynamic EAGLE simulation to determine the black hole content of the sample of simulated stripped nuclei from Paper I. We then compare this sample to the black holes in observed UCDs and simulated \hlcyan{galaxies}. Our aims are to determine whether the masses and numbers of black holes in stripped nuclei are consistent with observed UCDs known to contain black holes, whether the mass elevation ratio, $\Psi$ (\hlcyan{dynamical to stellar mass ratio}; see Section. \ref{section:elevmethodsml}), of the simulated stripped nuclei is consistent with observed UCDs, and whether supermassive black holes in UCDs are as common as those in normal galaxy nuclei.

When we refer to black holes throughout this paper, we are referring to central, primarily supermassive black holes (M\textsubscript{BH} $\gtrsim$ 10\textsuperscript{5}~\(\textup{M}_\odot\)) and not stellar mass black holes. When considering central black holes in lower mass objects such as globular clusters, we refer to the central black holes as massive or intermediate-mass black holes (M\textsubscript{BH} $\lesssim$ 10\textsuperscript{5}~\(\textup{M}_\odot\)). 

This paper is organized as follows. Section 2 briefly summarizes our stripped nuclei formation method from Paper I and how we determine further properties of that sample. The results of our research are presented in Section 3. We discuss the implications of our work for black holes in UCDs in Section 4, and a summary of our results is given in Section 5.

\label{sec:maths} 

\section{Methods}
\label{section:methods}
Here we summarize the stripped nuclei formation model of Paper I and describe how we determined the black hole masses of the simulated stripped nuclei.

\begin{table*}
	\centering
	\caption{\hl{Definitions of common terms and symbols used throughout this paper}}
	\label{tab:defs}
	\begin{tabular}{ll} 
		\hline
		Term &  \\
		\hline
            UCDs & Ultracompact dwarf galaxies, defined observationally as compact stellar systems with masses 2~$\times$~10\textsuperscript{6} to 10\textsuperscript{8}~\(\textup{M}_\odot\). UCD samples \\
             & may include both globular star clusters and the stripped nuclei of disrupted galaxies (see Sec.~\ref{section:intro}).\\
		Stripped nuclei & Surviving nuclei of disrupted galaxies. These may be observationally classified as either star clusters or galaxies (UCDs).\\
		Surviving galaxy & Galaxies in the EAGLE simulation that have not been \hlcyan{disrupted}, and survived to redshift z = 0\\
		Mass elevation & An increase in mass beyond that predicted from stellar population estimates measured by the mass elevation ratio.\\
		Mass elevation ratio & Ratio of the measured dynamical mass to the stellar population mass of an object, defined as $\Psi$\textsubscript{sim} or $\Psi$\textsubscript{obs}, depending on  \\
   & whether the object is simulated or observed.\\
		Galaxy mass & Mass of galaxies within the simulation. Refers either to surviving galaxies or galaxies prior to disruption depending on context\\
		Object mass & Masses of objects such as globular clusters, UCDs, or stripped nuclei depending on context.\\
         Total mass (M\textsubscript{tot}) & Total masses of objects such as globular clusters, UCDs, or stripped nuclei depending on context. Typically for stripped nuclei, \\
          & this is stellar mass+black hole mass since stripped nuclei are not expected to host much gas or dark matter\\

		\hline
	\end{tabular}
\end{table*}

\subsection{Creating a sample of simulated stripped nuclei}
\label{sec:sampcreation}
We used the hydrodynamical EAGLE simulations \citep{Crain_2015, Schaye2015} to create a sample of simulated stripped nuclei. Details of how the sample \hlcyan{was} created are given in Paper I and II. Here we briefly summarise the process. \hl{Readers should refer to those papers for greater detail.}

We used three sets of data from the EAGLE simulation: the online database of \citet{McAlpine2016}; the raw particle data, and a database we created that links the online database and the particle data.

To simulate the formation of stripped nuclei, we defined the central, most bound star particle (MBP) of a galaxy in the simulation \hlcyan{snapshot immediately prior to} merger as the nucleus of the galaxy and tracked this particle using the raw particle data. The online database was then used to determine the properties of the stripped nuclei, \hlcyan{using the properties of its progenitor galaxy from the snapshots before it was disrupted}.

A stripped nucleus is formed in a merger if the following conditions are satisfied:

\begin{enumerate}
  \item The \hlcyan{stellar} mass ratio between the merged galaxies is < 1/4. In a minor merger, only the less massive galaxy is disrupted and could therefore leave behind a stripped nucleus \citep{Qu2017}.  
  \item The stripped nucleus has been orbiting the more massive galaxy after the merger for a time shorter than its dynamical friction timescale, which is calculated using equation 7-26 from \citet{Binney1987}, modified with an eccentricity function as defined in Appendix B of \citet{Lacey1993}. 
\end{enumerate}
\hl{ Galaxies selected had stellar masses greater than \mbox{10\textsuperscript{7}~\(\textup{M}_\odot\)}, as this mass approaches the lower limit at which galaxies can be defined in the EAGLE simulation.}

\hl{At this limit, galaxies may be resolved by fewer than 10 stellar particles. However, orbits are determined by the dark matter halo, which is resolved by as many as \mbox{$\approx$}1000 particles at this mass. Note also that the number of low mass galaxies in EAGLE is calibrated to match observations \mbox{\citep{Schaye2015}}. For further discussion, see the Caveats section of Paper I.}

\hl{Note also that since the merger tree consists of 29 snapshots, the time at which galaxy mergers occur is not precisely resolved. The time between snapshots is \mbox{$\approx$0.5 Gyr at z $>\approx$ 1.5} (where most UCD formation is happening) and \mbox{$\approx$1 Gyr at z $<$ 0.5 \citep{McAlpine2016}}. Galaxies are disrupted on timescales of several Gyrs \mbox{\citep{Bekki2003, Pfeffer2013}}, and since we} \hlcyan{usually only} \hl{need to know whether a galaxy has been disrupted a timescale uncertainty of \mbox{$\pm$ 0.5 Gyr} is not significant. }

The \hlcyan{fraction} of nucleated galaxies was estimated from Figure 2 of \mbox{\citet{Sanchez2019}}, \hl{which used imaging data from the Next Generation Virgo Cluster Survey to study the properties of nuclei using 400 Virgo Cluster galaxies}. Nuclei mass estimates were based on Figure 9 of \mbox{\citet{Sanchez2019}}, which plots the ratio of nuclear star cluster to galaxy mass as a function of galaxy stellar mass. \hlcyan{ The stripped nucleus is tracked based on the location of the most bound particle at z=0.}


Galaxies in the EAGLE simulation may exhibit unstable behaviour during merger \citep{Qu2017}. Occasionally two merging galaxies will appear to swap masses between snapshots. This is due to the \textsc{subfind} \citep{Springel2001, Dolag2009} routine identifying particles attached to the central galaxy as instead being attached to the progenitor galaxy. 
Less than 5 per cent of merging galaxies exhibit this behaviour, but it can lead to an overestimation of stripped nuclei masses and numbers. 

Additionally, due to stripping during a merger event, a progenitor galaxy in the snapshot immediately before a merger may have lost stellar or black hole particles found in earlier snapshots. \hlcyan{Therefore,} for progenitor galaxies that do not exhibit switching behaviour, we take the maximum stellar and black hole masses in all snapshots before the merger. In galaxies \hlcyan{that} do switch masses, we take the stellar and black hole mass from the snapshot where the central galaxy stellar mass is at a maximum, which may lead to a slight underestimation of stellar and black hole mass for the progenitor galaxy in these cases.

\subsection{Black holes in the EAGLE simulation}
\label{sec:eaglebhs}


\hlcyan{The EAGLE simulation's model for black hole formation is described in \mbox{\citet{Schaye2015}}. Black hole formation is modelled by inserting a seed black hole in the centre of every halo with total mass greater than \mbox{10\textsuperscript{10}~\(\textup{M}_\odot\)} (corresponding to stellar mass \mbox{$\approx$ 10\textsuperscript{7}~\(\textup{M}_\odot\) -  10\textsuperscript{8}~\(\textup{M}_\odot\)} at high redshifts \mbox{\citep{McAlpine2016})} if it does not already contain a black hole, e.g. from a merger. This is done by converting the highest density gas particle into a collisionless black hole seed particle with subgrid mass \mbox{1.475~$\times$~10\textsuperscript{5}~\(\textup{M}_\odot\)}. The masses of the resulting supermassive black holes in EAGLE agree well with observations despite the challenges of modelling the process, as we discuss in Section \mbox{\ref{section:bhformation}}.}



The black hole masses given for each galaxy in the EAGLE database (table: SubHalo, column: BlackHoleMass) do not directly correspond to the mass of each galaxy's central supermassive black hole. Instead, they are a summed value of all the black holes assigned to that galaxy. Where the mass summation of all black holes assigned to the galaxy exceeds, M\textsubscript{BH} $=$ 10\textsuperscript{6}~\(\textup{M}_\odot\) this closely approximates the mass of the central supermassive black hole \citep{McAlpine2016}. 
Therefore when considering the sample of stripped nuclei that contain supermassive black holes, we will primarily focus on stripped nuclei with M\textsubscript{BH} $>$ 10\textsuperscript{6}~\(\textup{M}_\odot\). 

However, there is a growing amount of evidence that central black holes with M\textsubscript{BH} $<$ 10\textsuperscript{6}~\(\textup{M}_\odot\) exist in lower mass galaxies \citep[e.g.][]{Reines_2013, Brok_2015, Chilingarian_2018, Nguyen_2019}. The majority of stripped nuclei progenitor galaxies have stellar masses, M\textsubscript{*} $<$ 10\textsuperscript{10}~\(\textup{M}_\odot\), and a substantial amount of galaxies in this mass range are expected to contain supermassive black holes with M\textsubscript{BH} $<$ 10\textsuperscript{6}~\(\textup{M}_\odot\) \citep{Reines_2015}. For many UCDs, an elevated dynamical M/L ratio could be explained by a black hole with M\textsubscript{BH} $<$ 10\textsuperscript{6}~\(\textup{M}_\odot\) \citep{Voggel_2019}.
The black hole seed mass in EAGLE is approximately 1.475~$\times$~10\textsuperscript{5}~\(\textup{M}_\odot\), so we also consider stripped nuclei that contain a black hole mass more than twice this ($\approx$3~$\times$~10\textsuperscript{5}~\(\textup{M}_\odot\)) as a rule of thumb indicator of these stripped nuclei containing a supermassive black hole, due to the black hole showing evidence of growth via accretion of particles. \hlcyan{Note, however, that we do not consider these smaller black holes \mbox{(M\textsubscript{BH} $<$ 10\textsuperscript{6}~\(\textup{M}_\odot\))} to have reliable masses (Section \mbox{\ref{section:bhformation}}).}

Throughout this paper, we assume that every progenitor galaxy \hl{that hosts a single central supermassive black hole also has a nucleus (that the SMBH is found within) and thus can} become a stripped nucleus. We consider this a fair assumption because the growth of a nucleus and a black hole are linked, and nuclear star clusters and supermassive black holes coexist at lower galaxy masses \citep{Seth_2008}.  
If a disrupted galaxy does not have a nucleus, the disruption would likely produce a hyper-compact object, a supermassive black hole surrounded by a dense cluster of stars \citep{Merritt_2009, Leigh2013}. Since our objective is to understand the population of supermassive black holes in galaxy clusters that do not exist within galaxy nuclei, these black holes are worth including even if they do not exist within stripped nuclei.


\subsection{Modelling a sample of black holes in simulated stripped nuclei}
In Paper I, we created samples of simulated stripped nuclei in 7 massive EAGLE clusters. Here we predict the black hole content of the stripped nuclei by assuming that the black holes in their progenitor galaxies are unaffected by the stripping process. \hlcyan{The original mass of the black hole within the progenitor galaxy determines the black hole mass. As described in Section. \mbox{\ref{sec:sampcreation}}, we set the stripped nuclei's black hole masses as the progenitor galaxy's maximum black hole mass from all snapshots before it merges, unless switching behaviour occurs, in which case we take the stellar and black hole mass from the snapshot where the central galaxy stellar mass is at a maximum. }

This method gives us a more accurate measurement of the black hole masses than taking the most recent snapshot; however, it may present an issue when comparing the black holes in stripped nuclei to the black holes in surviving galaxies. The mass loss effect is not as strong for surviving galaxies as fewer will be experiencing mergers\hlcyan{. However, some} lower-mass surviving galaxies may have less reliable black hole masses.

One issue with including black holes in our model of stripped nucleus formation is that our calculation of nucleus mass is based purely on progenitor galaxy stellar mass. There is no input from other causal factors like the mass of the central supermassive black hole. In reality, the mass of the nucleus and its central supermassive black hole are linked because supermassive black hole growth is primarily fueled by the accretion of gas, which will also drive stellar formation \citep{Rafferty2006, Somerville2008, Storchi_Bergmann_2019}. 

However, when we calculate the stripped nucleus mass, we do not incorporate the black hole-nucleus mass relation. As a result, our model may produce stripped nuclei that contain supermassive black holes more massive than the nucleus, which we consider unphysical for lower-mass galaxies. To model the link between the nucleus and black hole mass, we assume in our results that a nucleus cannot host a black hole that makes up more than 30 per cent of its mass. This assumption is based on \citet{Mieske2013}'s estimate that observed UCDs host black holes that, on average, make-up 10-15 per cent of the mass of the nucleus, consistent with the expected black hole mass fractions of nuclear clusters in UCD progenitor galaxies. 

Confirmed supermassive black holes within UCDs have been found to make up 2-18 per cent of the mass of the UCD \citep{Seth2014,Ahn2017, Ahn2018, Afanasiev_2018}, so we consider the 30 per cent limit reasonable.
If, after the mass of the stripped nucleus has been calculated by the method described in Section \ref{sec:sampcreation}, we find the mass of the supermassive black hole exceeds 30 per cent of the mass of a stripped nucleus we repeatedly recalculate the stripped nucleus mass by the method described in Section. \ref{sec:sampcreation} until it is massive enough to host the supermassive black hole without the supermassive black hole exceeding 30 per cent of the total mass.

\subsection{Estimating dynamical mass-to-light ratios}
\label{section:methodsml}

The mass of a star cluster or compact galaxy can be estimated in different ways. One method estimates an object's mass by multiplying the object's luminosity by the stellar population mass-to-light ratio, which is calculated from the object's observed metallicity and age.

Alternatively, one can estimate an object's mass by using the velocity dispersion to calculate the mass via the virial theorem. When these two methods are compared, the velocity dispersion calculation of mass will appear inflated if a star cluster or UCD contains a central supermassive black hole. This is because a supermassive black hole within a compact object influences stellar velocities increasing the velocity dispersion, affecting the dynamical mass calculation. A central supermassive black hole will result in a velocity dispersion that is higher than the velocity dispersion of a star cluster with the same mass spread evenly throughout the object \citep{Merritt_2009}. An increase in dark mass affects the potential. From the virial theorem, an increase in dark mass in the centre lowers the potential more than the same mass spread out, so the average velocity increases. Typically the increase is 4-5 times higher than the same amount of mass distributed evenly \citep{Mieske2013}.

\subsubsection{Observability of supermassive black holes in UCDs}

The detectability of a supermassive black hole within a UCD depends on it making up a large enough mass fraction to be dynamically detectable from integrated spectra. The detectability of a black hole depends on its distance from Earth, its mass, the mass of the host UCD, and the ratio of the two. More massive black holes in more massive UCDs that make up a large mass fraction of the UCD are easier to detect due to having a stronger dynamical effect, better signal-to-noise, a larger absolute difference and larger velocity dispersion, respectively. The high-quality observations of \citet{Ahn2017}  were able to make a confirmed detection of a supermassive black hole in a UCD that makes up only $\approx$2 per cent of the UCD mass in a high mass UCD.

In our results, we define a black hole as being detectable if it makes up more than 3 per cent of a stripped nucleus, with this value chosen for its consistency with the observations. Our stripped nuclei sample will likely contain supermassive black holes that make up lower percentages of the stripped nuclei masses, but as observers have difficulty detecting these, we chose to exclude them when comparing our results to observed UCDs.

\subsubsection{Calculating the elevated mass ratios of stripped nuclei}
\label{section:elevmethodsml}

The exact influence that a supermassive black hole has on a stripped nucleus's observed dynamical mass can also be directly calculated from modelling/simulations. We have used the code described in \citet{Hilker2007} to set up a dynamical model of a stripped nucleus. The code works by the following steps:

\begin{enumerate}
  \item The parameters for the luminosity profile of the stripped nucleus are chosen. For our sample, we require the stripped nucleus mass, black hole mass and the S\'ersic index and the effective radius of the S\'ersic profile for a star cluster. The parameters we use for the sample are the effective radius, r\textsubscript{h} = 30 pc and S\'ersic index $n = 3.23 \pm 0.46$, with the S\'ersic index value chosen from the mean of 21 UCDs imaged by \citet{Evstig2008A} and uncertainties stemming from the standard error. 
  \item The 2-dimensional surface density profile as quantified by the above parameters is deprojected using Abel's integral equation \citep[equation 1B-59 of][]{Binney1987} into a 3-dimensional density profile $\rho(r)$.
  \item The 3-dimensional density profile is used to calculate the cumulative mass function $M(<r)$, the potential energy $\phi(r)$ with black holes of varying mass fraction, and the energy distribution function $f(E)$ using Eddington's method \citep[equation 4-140b of][]{Binney1987}.
  \item The cumulative mass function is normalised to 1. The function is numerically inverted and used to set up the central distance of each particle. The particle positions are then set up isotropically in space based on the observations that UCDs are near-isotropic \citep{Seth2014}. Our star cluster is set up with 500 000 particles.
  \item The potential energy of each particle is calculated for the chosen central distance. Each particle's velocity is calculated from the distribution function, $f(E)$. The three velocity components are then calculated, assuming that the velocities are isotropic. 
 \item The final result is a spherically symmetric N-body representation of the stripped nucleus.
\end{enumerate}

We use this model to directly calculate the mass elevation ratio ($\Psi$) of the simulated stripped nuclei as a function of the black hole mass fraction. The relevant parameters for our model are the S\'ersic index and the fraction between black hole mass and total mass. The input effective radius does not affect $\Psi$.

The model calculates the velocity of each particle, from which the \hlcyan{star cluster's }velocity dispersion, $\sigma$, can be calculated. We set up clusters with and without supermassive black holes, calculate their velocity dispersions, and from that, the mass elevation ratio $\Psi$\textsubscript{sim} = $\sigma$\textsubscript{BH}\textsuperscript{2}/$\sigma$\textsubscript{noBH}\textsuperscript{2} where $\sigma$\textsubscript{BH}\textsuperscript{2} and $\sigma$\textsubscript{noBH}\textsuperscript{2} are the velocity dispersions with and without a supermassive black hole, respectively. To exclude outliers, stars in the models containing black holes with higher velocities than the fastest-moving stars in the models without black holes are removed before \hlcyan{calculating} $\sigma$\textsubscript{BH}.

This modelled value of the mass elevation ratio can then be compared with observations of the dynamical and stellar population M/L ratios of observed UCDs, $\Psi$\textsubscript{obs} = (M/L)\textsubscript{dyn}/ (M/L)\textsubscript{pop}. Note that the observers' calculation of (M/L)\textsubscript{pop} is 
subject to modelling uncertainties such as the ages and metallicities of the stellar population, and the estimation of (M/L)\textsubscript{dyn} is subject to uncertainties in properties such as the measured velocity dispersion $\sigma$, the density profile, and the assumed distance to the UCD.  

\section{Results}

In this section, we present results from the analysis of the EAGLE simulations described in Section \ref{section:methods}, and compare our results with observations of UCDs. 

\subsection{Stripped nuclei black hole masses}
\label{section:bhm}

\begin{figure}
\centering
	\includegraphics[width=1.0\linewidth]{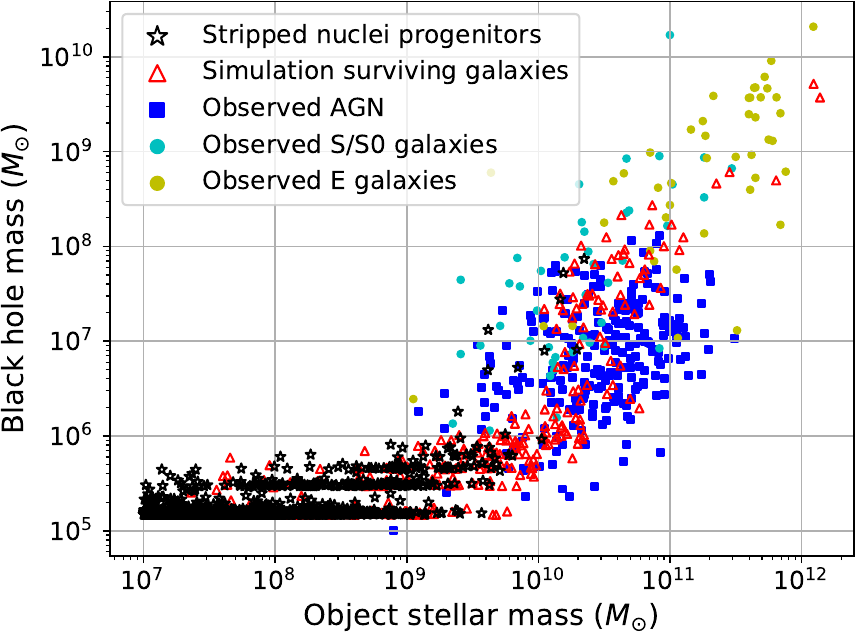}
    \caption[Stellar mass compared to black hole mass for observed and simulated galaxies]{Stellar mass compared to black hole mass for observed galaxies and simulated galaxies, measured at redshift z = 0 including the progenitor galaxies of stripped nuclei, measured at merger time. The observed AGNs are from \citet{Reines_2015}, and the other observed galaxies are from \citet{Kormendy2013}.}
    \label{fig:obssim}
\end{figure}

Figure~\ref{fig:obssim} plots calibrated z = 0 black hole masses against stellar masses for stripped nuclei progenitor galaxies and surviving galaxies in the most massive cluster in the EAGLE simulation, along with observed galaxies from \citet{Kormendy2013} and \citet{Reines_2015}. This figure verifies the result in \citet{Schaye2015} that the black holes in simulated galaxies are consistent with \hlcyan{those} in observed galaxies, although the simulated relation exhibits less scatter. The stripped nuclei progenitor galaxies are also consistent with this relation, although the progenitor galaxies with stellar mass 10\textsuperscript{9}~\(\textup{M}_\odot\)-10\textsuperscript{10}~\(\textup{M}_\odot\) have slightly more massive black holes than the surviving simulated galaxies. This may be because the masses of observed and simulated galaxies are taken at redshift z = 0, while stripped nuclei progenitor galaxies form and merge in the \hlcyan{universe's early stages}. The stripped nuclei progenitors would, therefore, lie above the relation since black hole growth may outpace stellar mass growth for early galaxies \citep[e.g.][]{Merloni2010, Willott_2013, Decarli_2018}.

\begin{figure}
\centering
	\includegraphics[width=1.0\linewidth]{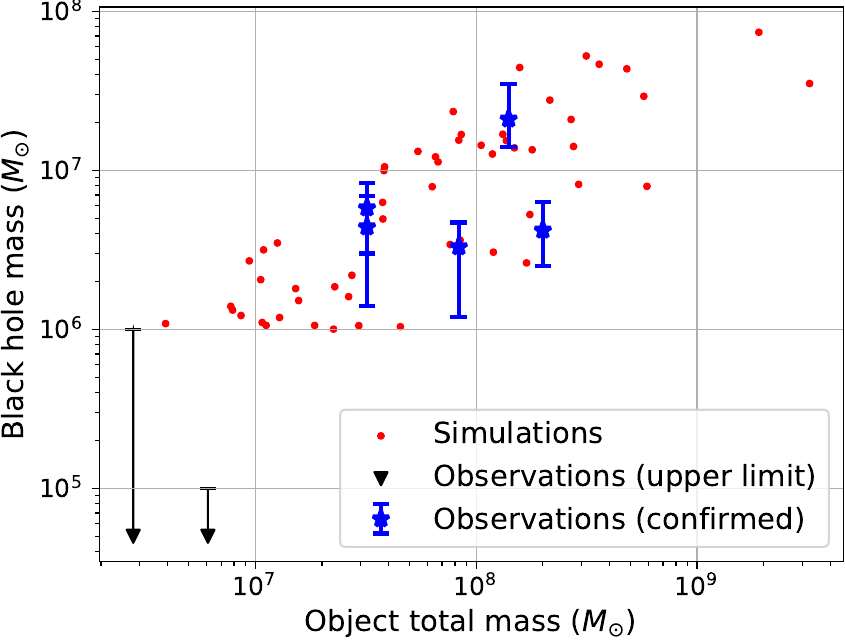}
    \caption[Black hole mass compared to the total masses of simulated stripped nuclei and observed UCDs.]{Black hole mass compared to \hl{the total masses of} simulated stripped nuclei and observed UCDs. The observed UCDs with confirmed black holes are from the studies by \citet{Seth2014}, \citet{Ahn2017}, \citet{Ahn2018}, and \citet{Afanasiev_2018}. Also included are two UCDs with $3\sigma$ upper limits on black hole masses from \citet{Voggel_2018}.}
    \label{fig:bhmasses}
\end{figure}

Figure~\ref{fig:bhmasses} plots derived black hole masses against derived total masses for simulated stripped nuclei with black hole masses above 10\textsuperscript{6}~\(\textup{M}_\odot\) for the seven massive clusters in EAGLE and UCDs whose black hole content has been studied. The 5 UCDs that are confirmed to contain black holes are M60-UCD1 \citep{Seth2014}, VUCD3, M59cO \citep{Ahn2017}, M59-UCD3 \citep{Ahn2018} and UCD3 \citep{Afanasiev_2018}. Also included are UCD 330 and UCD 320 \citep{Voggel_2018}, for which only upper limits on black hole masses have been found. Figure~\ref{fig:bhmasses} shows that the black hole masses of the simulated stripped nuclei are consistent with the black hole masses of observed UCDs. 

The masses and mass fractions of these black holes in UCDs are a reliable indicator of them being stripped nuclei. This result is a strong test of our model of UCD formation since the lack of a relation between nucleus mass and black hole mass in our model means that it is not a guarantee that our model would match observations. It is also a weak test of the EAGLE simulations since \hl{black holes in the simulations are calibrated to match low-redshift galaxy properties and the effect of black hole feedback by calibrating the black hole mass-stellar mass relation, and not the masses of black holes at the high redshifts that stripped nuclei form} \citep{Mayes2021}.


The most massive black hole we find in a stripped nucleus in any cluster has a mass of 7~$\times$~10\textsuperscript{7}~\(\textup{M}_\odot\). However, this object has a progenitor galaxy mass of 6~$\times$~10\textsuperscript{10}~\(\textup{M}_\odot\) and a predicted nucleus mass of 3~$\times$~10\textsuperscript{9}~\(\textup{M}_\odot\), which could mean that it may more closely resemble a compact elliptical galaxy than a UCD. Compact ellipticals have masses $>$~10\textsuperscript{9}~\(\textup{M}_\odot\) and $R$\textsubscript{eff} > 100 pc while observed UCDs typically have maximum masses of a few times 10\textsuperscript{8}~\(\textup{M}_\odot\) and $R$\textsubscript{eff} < 100 pc \citep[e.g.][]{Mieske2006}. 
Excluding this object, the most massive black hole in a stripped nucleus for the seven clusters ranges from 1.2~$\times$~10\textsuperscript{7}~\(\textup{M}_\odot\) to 5~$\times$~10\textsuperscript{7}~\(\textup{M}_\odot\). This range is consistent with the most massive black hole found in a UCD, M60-UCD1, which has a mass of 2~$\times$~10\textsuperscript{7}~\(\textup{M}_\odot\) and is found in the massive Virgo cluster. Table~\ref{tab:massbhs} lists the most massive black hole in the stripped nuclei of each simulated cluster. The most massive black hole mass increases with cluster mass, although there is a reasonable amount of scatter.

\begin{table*}
	\centering
	\caption{The most massive black hole in the stripped nuclei population of each EAGLE cluster}
	\label{tab:massbhs}
	\begin{tabular}{lcc} 
		\hline
		Cluster-ID & Cluster Friend-of-Friends Mass (\(\textup{M}_\odot\)) & Most massive black hole mass (\(\textup{M}_\odot\))\\
		\hline
		28000000000000 & 6.42~$\times$~10\textsuperscript{14} & 7.4~$\times$~10\textsuperscript{7}\\
		28000000000001 & 6.22 $\times$ 10\textsuperscript{14} & 4.7~$\times$~10\textsuperscript{7}\\
		28000000000002 & 3.76 $\times$ 10\textsuperscript{14} & 4.4~$\times$~10\textsuperscript{7}\\
		28000000000003 & 3.48 $\times$ 10\textsuperscript{14} & 1.4~$\times$~10\textsuperscript{7}\\
		28000000000004 & 2.50 $\times$ 10\textsuperscript{14} & 1.2~$\times$~10\textsuperscript{7}\\
		28000000000005 & 2.31 $\times$ 10\textsuperscript{14} & 4.4~$\times$~10\textsuperscript{7}\\
		28000000000006 & 2.05 $\times$ 10\textsuperscript{14} & 2.0~$\times$~10\textsuperscript{7}\\
		\hline
	\end{tabular}
\end{table*}

\subsection{What percentage of stripped nuclei contain black holes?}
\label{section:lowmass}

In Paper I, we found that the most massive cluster in the EAGLE simulation contains approximately 2182 stripped nuclei. Of those 400 $\pm$ 50 are more massive than 2~$\times$~10\textsuperscript{6}~\(\textup{M}_\odot\), 51.8 $\pm$ 7.7 are more massive than 10\textsuperscript{7}~\(\textup{M}_\odot\), and 2.3 $\pm$ 1.1 are more massive than 10\textsuperscript{8}~\(\textup{M}_\odot\). 

Here we find that the population of stripped nuclei in this cluster contains 10 black holes above a mass of 10\textsuperscript{6}~\(\textup{M}_\odot\). Assuming all stripped nuclei that host M $>$ 10\textsuperscript{6}~\(\textup{M}_\odot\) black holes are more massive than M\textsubscript{tot} = 10\textsuperscript{7}~\(\textup{M}_\odot\), this represents a high mass stripped nuclei black hole occupation fraction of $\approx$20 per cent for this massive cluster and a 0 per cent occupation rate for low mass stripped nuclei with M\textsubscript{tot} $<$ 10\textsuperscript{7}~\(\textup{M}_\odot\). More likely, the occupation fraction for high-mass stripped nuclei would be slightly below 20 per cent, and a small percentage of low-mass stripped nuclei would also contain black holes. The other 6 clusters in EAGLE show similar results. Table~\ref{tab:occupfrac} shows the number of M\textsubscript{tot} $>$ 10\textsuperscript{7}~\(\textup{M}_\odot\) stripped nuclei,  M\textsubscript{BH} $>$ 10\textsuperscript{6}~\(\textup{M}_\odot\) black holes and the occupation fractions. The high mass stripped nuclei black hole occupation fractions range from 15 to 43 per cent with an overall mean of 27 per cent.

\begin{table*}
	\centering
	\caption{Occupation fraction of  M\textsubscript{BH} $>$ 10\textsuperscript{6} \(\textup{M}_\odot\) black holes in M\textsubscript{tot} $>$ 10\textsuperscript{7} \(\textup{M}_\odot\) stripped nuclei}
	\label{tab:occupfrac}
	\begin{tabular}{lccc} 
		\hline
		Cluster ID & Number of M\textsubscript{tot} $>$  10\textsuperscript{7} \(\textup{M}_\odot\) & Number of  M\textsubscript{BH} $>$ 10\textsuperscript{6} \(\textup{M}_\odot\) & Occupation fraction\\
		  & Stripped nuclei & Black holes &  \\
		\hline
		28000000000000 & 51.8 & 10 & 19\%\\
		28000000000001 & 52.2 & 14 & 27\%\\
		28000000000002 & 38.6 & 8 & 21\%\\
		28000000000003 & 20.6 & 7 & 34\%\\
		28000000000004 & 19.4 & 3 & 15\%\\
		28000000000005 & 18.6 & 4 & 27\%\\
		28000000000006 & 18.8 & 8 & 43\%\\
		\hline
	\end{tabular}
\end{table*}

However, this calculation represents a bare minimum occupation fraction, based on the unreliability of EAGLE black hole masses below
10\textsuperscript{6}~\(\textup{M}_\odot\) (due to the closeness to the seed mass). Extending the comparison to consider stripped nuclei containing a black hole with a mass more than twice the seed mass ($\approx$3~$\times$~10\textsuperscript{5}~\(\textup{M}_\odot\)), we find that 213 stripped nuclei in the most massive cluster contain supermassive black holes. Assuming higher-mass stripped nuclei are more likely to contain supermassive black holes, this represents an occupation fraction of 53 per cent for stripped nuclei above a total mass of 2~$\times$~10\textsuperscript{6}~\(\textup{M}_\odot\) and is a likely indicator that all stripped nuclei with total mass $>$~10\textsuperscript{7}~\(\textup{M}_\odot\) contain supermassive black holes. Table~\ref{tab:occupfracv2} depicts the black hole occupation fractions of stripped nuclei with M\textsubscript{BH} $>$ 3~$\times$~10\textsuperscript{5}~\(\textup{M}_\odot\) for all 7 EAGLE clusters, ranging from 41 to 58 per cent with a mean of 51 per cent.  

\begin{table*}
	\centering
	\caption{Occupation fraction of  M\textsubscript{BH} $>$3~$\times$~10\textsuperscript{5} \(\textup{M}_\odot\) black holes in M\textsubscript{tot} $>$  2~$\times$~10\textsuperscript{6} \(\textup{M}_\odot\) stripped nuclei}
	\label{tab:occupfracv2}
	\begin{tabular}{lccc} 
		\hline
		Cluster ID & Number of M\textsubscript{tot} $>$ 2~$\times$~10\textsuperscript{6} \(\textup{M}_\odot\) & Number of  M\textsubscript{BH} $>$ 3~$\times$~10\textsuperscript{5} \(\textup{M}_\odot\) & Occupation fraction\\
		  & Stripped nuclei & Black holes &  \\
		\hline
		28000000000000 & 400 & 213 & 53\%\\
		28000000000001 & 384.3 & 190 & 49\%\\
		28000000000002 & 256.4 & 129 & 50\%\\
		28000000000003 & 211.9 & 105 & 50\%\\
		28000000000004 & 137.9 & 73 & 53\%\\
		28000000000005 & 164.8 & 67 & 41\%\\
		28000000000006 & 112.9 & 65 & 58\%\\
		\hline
	\end{tabular}
\end{table*}


In Paper I, we found a number of stripped nuclei with M\textsubscript{tot} $<$ 2~$\times$~10\textsuperscript{6}~\(\textup{M}_\odot\) that may be observed as globular clusters and not UCDs, including a substantial number with masses between 10\textsuperscript{4}~\(\textup{M}_\odot\) and 10\textsuperscript{6}~\(\textup{M}_\odot\) that would be more similar to globular clusters than UCDs purely on a mass basis. Here we consider the possibility that those low-mass stripped nuclei contain central supermassive black holes.

Considering the stripped nuclei with M\textsubscript{tot} $<$ 2~$\times$~10\textsuperscript{6}~\(\textup{M}_\odot\) that also have black holes, we find 71 stripped nuclei with black holes with mass $>$ 3~$\times$~10\textsuperscript{5}~\(\textup{M}_\odot\), 33 of which also have M\textsubscript{tot} $<$ 10\textsuperscript{6}~\(\textup{M}_\odot\). We find a total of 1780 $\pm$ 240 low-mass (M\textsubscript{tot} $<$ 2~$\times$~10\textsuperscript{6}~\(\textup{M}_\odot\)) stripped nuclei in this cluster. Of these, 1400 $\pm$ 190 have M\textsubscript{tot} $<$ 10\textsuperscript{6}~\(\textup{M}_\odot\). This would indicate that approximately 4 per cent of stripped nuclei with M\textsubscript{tot} $<$ 2~$\times$~10\textsuperscript{6}~\(\textup{M}_\odot\) and 2 per cent of stripped nuclei with M\textsubscript{tot} $<$ 10\textsuperscript{6}~\(\textup{M}_\odot\) could host massive black holes. 

The Virgo cluster is known to host 67300 $\pm$ 1440 globular clusters \citep{Durrell2014}. If the low mass stripped nuclei that contain black holes are observed as globular clusters, possibly $\approx$0.1 per cent of the globular cluster population in massive galaxy clusters host black holes that originated in galaxy nuclei.

However, it is important to note that the calculation of stripped nucleus mass is made purely based on the stellar mass of the host galaxy. Since supermassive black hole growth and nucleus growth are both correlated with stellar mass, the scatter in the calculation could easily mean that those stripped nuclei with low masses that contain black holes have true masses $>$ 2~$\times$~10\textsuperscript{6}~\(\textup{M}_\odot\) and would therefore be considered UCDs, not globular clusters. Additionally, even if they have M\textsubscript{tot} $<$ 2~$\times$~10\textsuperscript{6}~\(\textup{M}_\odot\), they may still be observed as UCDs and not globular clusters depending on other factors such as radii. For example, \citet{Forbes2013} find several low luminosity UCDs with sizes up to 40~pc and M\textsubscript{V}~$\sim  -8$ to $-9$ (M~$\sim$~few~$\times$~10\textsuperscript{5}~\(\textup{M}_\odot\)) that would be similar to these low mass stripped nuclei. Finally, we note that EAGLE black holes below 10\textsuperscript{6}~\(\textup{M}_\odot\) are also unreliable, and many of the black holes in low mass stripped nuclei we find fall very close to our lower limit of 3~$\times$~10\textsuperscript{5}~\(\textup{M}_\odot\). 

Because of this, the predicted number of stripped nuclei that may be observed as globular clusters that host black holes is more likely to be an upper limit. Our results are consistent with the low mass population of stripped nuclei containing no supermassive black holes. Hence, while many low-mass stripped nuclei may be observed as part of the globular cluster population, they may be indistinguishable from genuine globular clusters with regard to black hole content.


\subsection{Elevated dynamical M/L ratios of stripped nuclei}
\label{section:reselevml}
\subsubsection{The percentage of stripped nuclei with observable elevated dynamical masses}

\begin{figure}
\centering
	\includegraphics[width=1.0\linewidth]{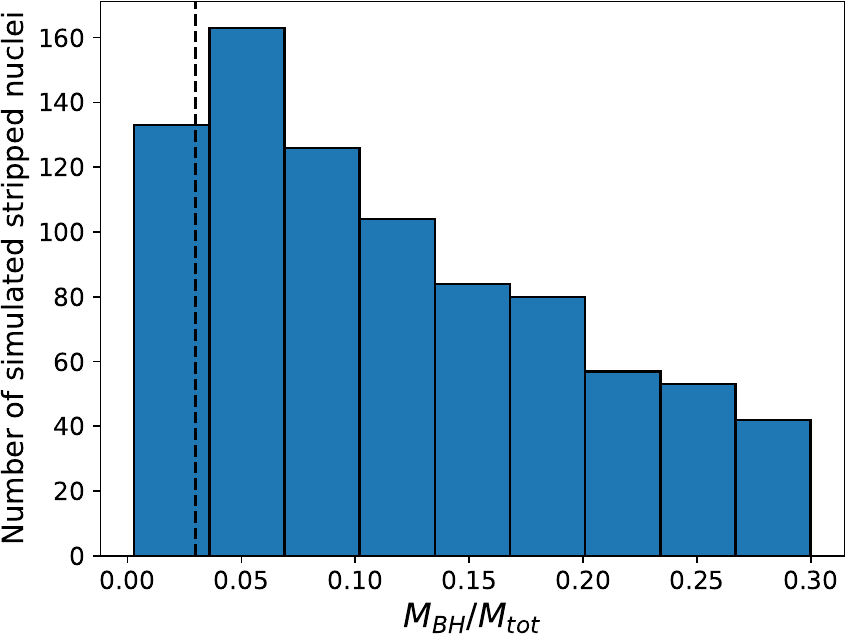}
    \caption[Distribution of the ratio of black hole mass to total mass for stripped nuclei]{Distribution of the ratio of black hole mass to total mass for stripped nuclei with M\textsubscript{BH} $>$ 3~$\times$~10\textsuperscript{5}~\(\textup{M}_\odot\). The dashed line shows the point above which black holes are likely dynamically detectable under the best observing conditions. \hl{Note that the masses of \mbox{M\textsubscript{BH} $<$ 10\textsuperscript{6}~\(\textup{M}_\odot\)} black holes are inherently unreliable; however the results are similar with \mbox{M\textsubscript{BH} $>$ 10\textsuperscript{6}~\(\textup{M}_\odot\)} black holes.}}
    \label{fig:ratio}
\end{figure}
As described in Section ~\ref{section:methodsml}, we define black holes in stripped nuclei as being dynamically detectable, under the best observing conditions, if they make up more than 3 per cent of the stripped nucleus's mass. We find 82 per cent of all stripped nuclei with mass $>$ 2~$\times$~10\textsuperscript{6}~\(\textup{M}_\odot\) contain black holes with M\textsubscript{BH} $>$ 3~$\times$~10\textsuperscript{5}~\(\textup{M}_\odot\) that meet this criterion. \hl{Note, however, that the masses of \mbox{M\textsubscript{BH} $<$ 10\textsuperscript{6}~\(\textup{M}_\odot\)} black holes are inherently unreliable.} When we consider only stripped nuclei with reliable black holes with  M\textsubscript{BH} $>$ 10\textsuperscript{6}~\(\textup{M}_\odot\), we find a similar fraction of 89 per cent. Figure~\ref{fig:ratio} plots the ratio for stripped nuclei with M\textsubscript{BH} $>$ 3~$\times$~10\textsuperscript{5}~\(\textup{M}_\odot\).

The overall conclusion is that a high percentage, 80-90 per cent, of supermassive or intermediate-mass black holes in stripped nuclei should be dynamically detectable. This result also predicts that a smaller percentage of stripped nuclei \hlcyan{should} contain supermassive black holes, which would not have detectable elevated dynamical masses. 

\subsubsection{Mass elevation ratios of the stripped nuclei population}
\begin{figure}
 \centering
 	\includegraphics[width=1.0\linewidth]{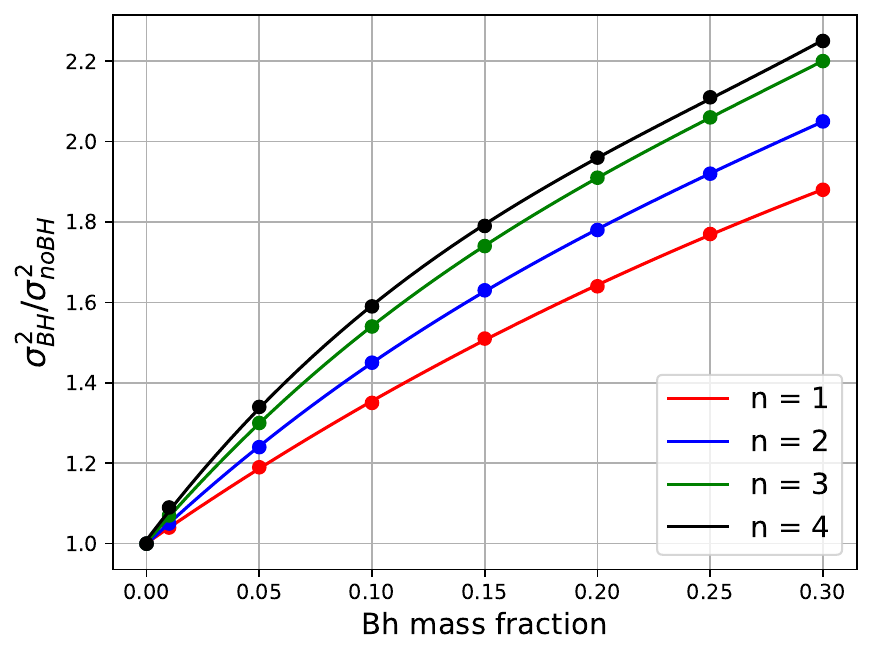}
     \caption[ML]{The mass elevation ratio, $\Psi$\textsubscript{sim}, of star clusters with varying black hole mass fraction and  S\'ersic index n. The elevation ratio is \hlcyan{calculated} with a dynamical model as described in Section \ref{section:methodsml}.} 
     \label{fig:masselev}
 \end{figure}

In Section~\ref{section:methodsml}, we described how the mass elevation ratio of star clusters can be calculated by modelling a star cluster with and without a supermassive black hole. Here we use this model to calculate the mass elevation ratios of our population of stripped nuclei. Figure~\ref{fig:masselev} plots the mass elevation ratio, $\Psi$\textsubscript{sim} for varying S\'ersic index n and black hole mass fraction. We found that \hlcyan{the} black hole mass fraction and S\'ersic index were \hlcyan{the} only relevant parameters when calculating $\Psi$\textsubscript{sim} and the mass of the star cluster was not a factor.

Since black hole masses M\textsubscript{BH} $<$ 10\textsuperscript{6}~\(\textup{M}_\odot\) are unreliable, accurately calculating the mean mass elevation ratio, $\Psi$, for the stripped nuclei population presents some difficulties. In particular, the lower limit of M\textsubscript{BH} = 3~$\times$~10\textsuperscript{5}~\(\textup{M}_\odot\) fails to consider the possibility of low-mass stripped nuclei containing lower-mass black holes than the limit. This will inflate the mean mass elevation ratio of the stripped nuclei population. Modelling with observed correlations between galaxy stellar mass and black hole mass may help this issue, but as UCDs form and merge at high redshifts, this would require high-redshift observations.

Using the sample with only reliable black holes with  M\textsubscript{BH} $>$ 10\textsuperscript{6}~\(\textup{M}_\odot\) will account for this issue but cause a different problem. Because only $\approx$27 per cent of high mass UCDs can contain black holes of this mass, the mean mass elevation ratio calculation of the population of stripped nuclei will appear over-inflated without the inclusion of the objects with lower mass black holes.

\begin{figure}
\centering
	\includegraphics[width=1.0\linewidth]{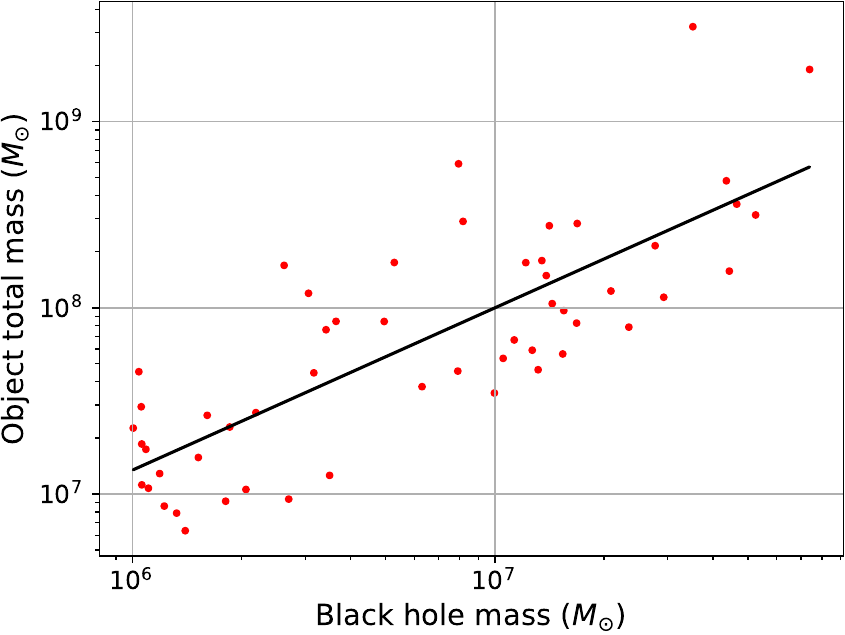}
    \caption[Total mass compared to black hole mass for simulated stripped nuclei.]{Total mass compared to black hole mass for simulated stripped nuclei with a line of best fit. The plot is made with only the reliable black holes with  M\textsubscript{BH} $>$ 10\textsuperscript{6}~\(\textup{M}_\odot\). We use the line of best fit to extrapolate the relationship to lower black hole masses. By this method, we can correct the  M\textsubscript{BH}/M\textsubscript{tot} relationship for the incompleteness of black holes with mass  M\textsubscript{BH} $<$ 10\textsuperscript{6}~\(\textup{M}_\odot\)}
    \label{fig:bhmtotmasslobf}.
\end{figure}

We can correct the mean mass elevation for this issue by extrapolating the mean relation between total mass and black hole mass to lower masses as follows.
\begin{itemize}
\item We first plot total mass as a function of black hole mass for our sample of simulated stripped nuclei in Figure~\ref{fig:bhmtotmasslobf} and apply a line of best fit to the plot. This plot allows us to extrapolate the total mass-black hole mass relationship to the lower range of black hole masses. 
\item From this plot, we can determine the mean total mass of the stripped nuclei population for a given black hole mass. In turn, we can extrapolate the black hole mass fraction down to lower black hole masses than what can be accurately calculated from the simulation.
\item We find the data has the relation $\log \textsubscript{10}~M\textsubscript{tot}~=~(0.92 \pm 0.09)~\log\textsubscript{10}~M\textsubscript{BH}~+(1.58 \pm 0.61)$ from Figure~\ref{fig:bhmtotmasslobf}. 
\item We next use this equation to calculate the mean expected black hole mass for each stripped nucleus with M\textsubscript{tot} $>$ 10\textsuperscript{7}~\(\textup{M}_\odot\). We find this sample has mean  M\textsubscript{BH} = 5.46~$\times$~10\textsuperscript{6}~\(\textup{M}_\odot\) and M\textsubscript{tot} $=$ 6.00~$\times$~10\textsuperscript{7}~\(\textup{M}_\odot\). Therefore, the mean black hole mass fraction for stripped nuclei with M\textsubscript{tot} $>$ 10\textsuperscript{7}~\(\textup{M}_\odot\) is $0.09 \pm 0.01$ per cent. Calculating mass elevation ratio using S\'ersic index $n = 3.23 \pm 0.46$, from \citet{Evstig2008A} we find $\Psi$\textsubscript{sim} $= 1.51^{+0.06}_{-0.04}$ for the population of stripped nuclei with M\textsubscript{tot} $>$ 10\textsuperscript{7}~\(\textup{M}_\odot\). 
\item We find $\Psi$\textsubscript{sim} increases for higher mass objects. Table~\ref{tab:masselev} shows the mass elevation ratio at different total masses and the mean mass elevation ratio for high-mass stripped nuclei.
\item \hl{This calculation assumes that most, or all stripped nuclei with \mbox{M $>$ 10\textsuperscript{7}~\(\textup{M}_\odot\)} contain massive black holes.}
\end{itemize}

\begin{figure}
\centering
	\includegraphics[width=1.0\linewidth]{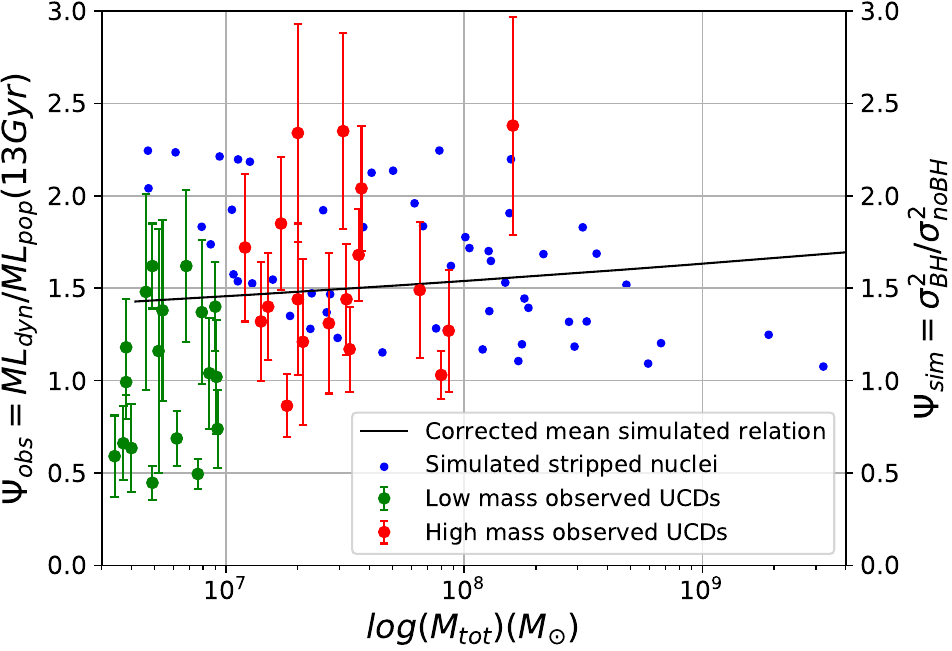}
    \caption[$\Psi$ for simulated stripped nuclei and observed UCDs]{Mass elevation ratio, $\Psi$\textsubscript{sim} for simulated stripped nuclei with  M\textsubscript{BH} $>$ 10\textsuperscript{6}~\(\textup{M}_\odot\). The black line represents the mean mass elevation ratio of the stripped nuclei population, using S\'ersic index $n = 3.23$ corrected for the incomplete sampling of low-mass black holes. 
    Also included are observations of $\Psi$\textsubscript{obs} for UCDs from \citet{Mieske2013}. 
    }
    \label{fig:mieske}
\end{figure}

\begin{table}
	\centering
	\caption{Mass elevation ratio of stripped nuclei that contain supermassive black holes}
	\label{tab:masselev}
	\begin{tabular}{lc} 
		\hline
		Stripped nuclei mass, M\textsubscript{tot}  & $\Psi$\textsubscript{sim} \\
		
		\hline
		M\textsubscript{tot} =  10\textsuperscript{7}~\(\textup{M}_\odot\) & $1.46^{+0.04}_{-0.01}$ \\
		M\textsubscript{tot} =  10\textsuperscript{8}~\(\textup{M}_\odot\) & $1.53^{+0.04}_{-0.01}$ \\
		\hline
		Stripped nuclei mass, M\textsubscript{tot}  & Mean $\Psi$\textsubscript{sim} \\
		\hline
		  M\textsubscript{tot} $>$ 10\textsuperscript{7}~\(\textup{M}_\odot\)  & $1.51^{+0.06}_{-0.04}$\\
		\hline
	\end{tabular}
\end{table}
Figure~\ref{fig:mieske} plots $\Psi$\textsubscript{sim} for simulated stripped nuclei with  M\textsubscript{BH} $>$ 10\textsuperscript{6}~\(\textup{M}_\odot\). The black line is the mean mass elevation, corrected as described above. Also included are observations of $\Psi$\textsubscript{obs} for UCDs from \citet{Mieske2013}. Two different y-axes are used for the observed and simulated objects.

Below M\textsubscript{tot} $\approx$ 5~$\times$~10\textsuperscript{7}~\(\textup{M}_\odot\), the mass elevation ratios of the simulated stripped nuclei appear to increase. This roughly corresponds to the mass below which we expect to find  M\textsubscript{BH} $<$ 10\textsuperscript{6}~\(\textup{M}_\odot\) black holes in Figure~\ref{fig:bhmtotmasslobf}. As a result, this increase is artificial and should not be interpreted as the mass elevation ratios of stripped nuclei increasing towards lower masses.

The mass elevation ratio of stripped nuclei also appears to decline for higher masses. However, caution must also be used when comparing these objects to observed UCDs. At high masses, these stripped nuclei may no longer resemble UCDs but instead may be considered compact elliptical galaxies. Compact ellipticals have stellar masses of 10\textsuperscript{8}~\(\textup{M}_\odot\) $<$ M\textsubscript{*} $<$ 10\textsuperscript{10}~\(\textup{M}_\odot\) and galaxy sizes between 100 - 900 pc. They may also form by tidal stripping \citep[e.g.][]{Bekki2003, Graham_2003, Ferr_Mateu_2021b}. Estimating the true mass elevation ratio for objects at these masses would require a complete study of these objects using the EAGLE simulation. Our sample of simulated stripped nuclei does not include all of these objects, as, unlike the stripped nuclei, they may retain enough material to be still considered galaxies by EAGLE. As such, the mean mass elevation ratio of the population of objects above M\textsubscript{tot} $\approx$ a few times 10\textsuperscript{8}~\(\textup{M}_\odot\) is unreliable. (The highest mass UCD from \citet{Mieske2013} has mass $M$ $\approx$  2~$\times$~10\textsuperscript{8}~\(\textup{M}_\odot\)).

\subsection{Black hole growth during merger}
It is possible that during the merger process, the black hole experiences growth due to gas driven to the centre of a galaxy feeding the black hole in a way that is not accurately modelled by the limited resolution of the simulations \citep{Ricarte2020}. The exact growth this may cause is unknown, but it could potentially double the masses of the black holes as they currently are in the simulation. This would increase the number of black holes in stripped nuclei with  M\textsubscript{BH} $>$ 10\textsuperscript{6}~\(\textup{M}_\odot\) and possibly raise the percentage of stripped nuclei with elevated dynamical masses. In the most massive cluster, we find 41 stripped nuclei with M\textsubscript{BH} $>$ 5~$\times$~10\textsuperscript{5}~\(\textup{M}_\odot\), so if gas-fueled growth during the merger is a factor, we would expect $\approx$80 per cent of massive UCDs to host black holes with  M\textsubscript{BH} $>$ 10\textsuperscript{6}~\(\textup{M}_\odot\). This could affect the prediction of the number of stripped nuclei with dynamically detectable black holes and the mass elevation ratios of stripped nuclei. However, gas-fueled growth during the merger would also cause the nucleus to grow, complicating these predictions. This result is also dependent on the unreliable masses of black holes in the EAGLE simulation below 10\textsuperscript{6}~\(\textup{M}_\odot\), and the unknown rate of growth a black hole and nucleus could experience during a merger.

\subsection{Comparing black holes within stripped nuclei to black holes within surviving galaxy nuclei}
\label{section:nuclei}



\begin{figure*}
\begin{multicols}{2}
    \includegraphics[width=\linewidth]{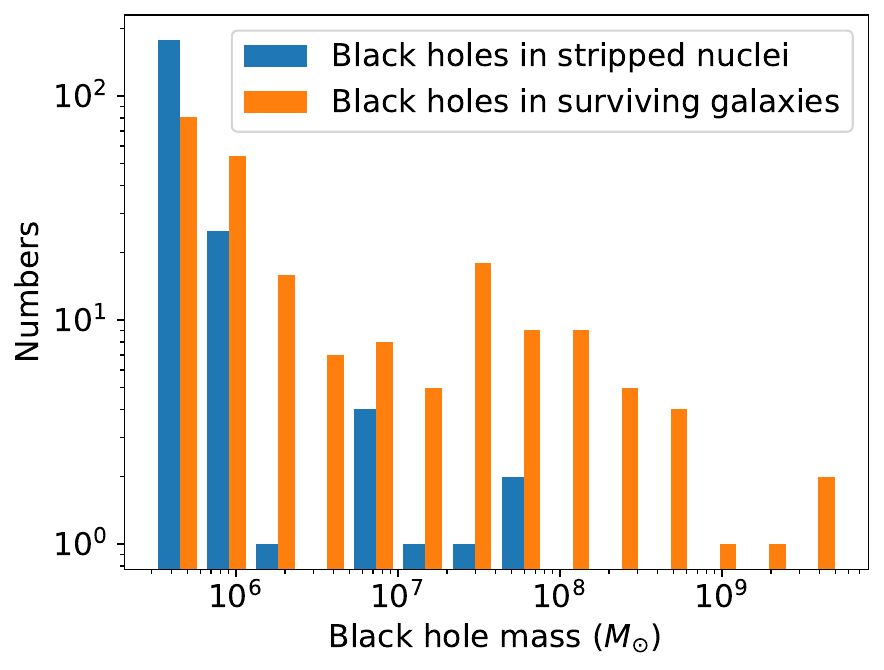}\par 
    \includegraphics[width=\linewidth]{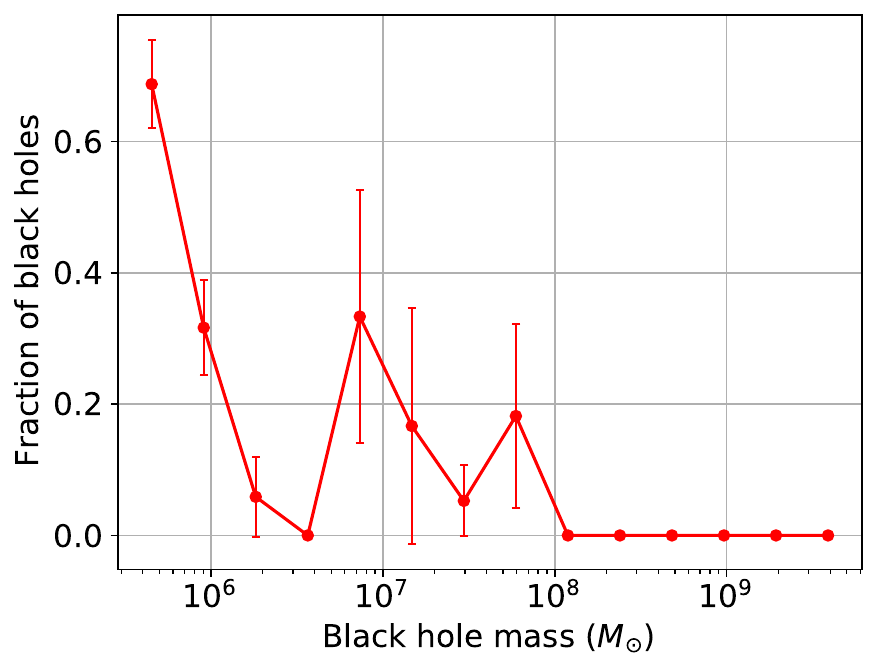}\par 
\end{multicols}
\caption[The number of supermassive black holes in progenitor and surviving galaxies.]{A comparison of the black hole population within stripped nuclei to the black hole population within surviving galaxies. The comparison is made at a given black hole mass for a single simulated cluster. \textbf{Left:} A comparison of the number of supermassive black holes in stripped nuclei with those in surviving galaxies. \textbf{Right:} The fraction of the black hole population that is found in stripped nuclei, defined as $f_{BHDis} = N_{BHDis}/(N_{BHDis}+N_{BHSur})$, with uncertainties \hlcyan{propagated as} \mbox{$\sqrt{N}$}}.
 \label{fig:bhsurprog}
\end{figure*}

Figure~\ref{fig:bhsurprog} plots the number of black holes with  M\textsubscript{BH} $>$ 3~$\times$~10\textsuperscript{5}~\(\textup{M}_\odot\) in the most massive cluster for stripped nuclei progenitor galaxies and surviving galaxies. We find that the most massive cluster in EAGLE contains 95 black holes more massive than 10\textsuperscript{6}~\(\textup{M}_\odot\), and 220 more massive than 3~$\times$~10\textsuperscript{5}~\(\textup{M}_\odot\). By comparison, this cluster contains 10 black holes in stripped nuclei with M\textsubscript{BH} $>$ 10\textsuperscript{6}~\(\textup{M}_\odot\) and 213 with M\textsubscript{BH} $>$ 3~$\times$~10\textsuperscript{5}~\(\textup{M}_\odot\), a ratio of black holes in stripped nuclei to galaxy nuclei of 0.1 and 0.97 respectively. This result indicates that while stripped nuclei may not contribute in large numbers to the population of the most massive black holes, they could rival or even outnumber the amount of lower-mass supermassive black holes and intermediate-mass black holes in galaxy nuclei. Table \ref{tab:surviv} shows the ratio of black holes in stripped nuclei to galaxy nuclei in the seven massive EAGLE clusters. The mean ratio for black holes with M\textsubscript{BH} $>$ 10\textsuperscript{6}~\(\textup{M}_\odot\) is 0.14, and the mean ratio for black holes with M\textsubscript{BH} $>$ 3~$\times$~10\textsuperscript{5}~\(\textup{M}_\odot\) is 0.85.

\begin{figure*}
\begin{multicols}{2}
    \includegraphics[width=\linewidth]{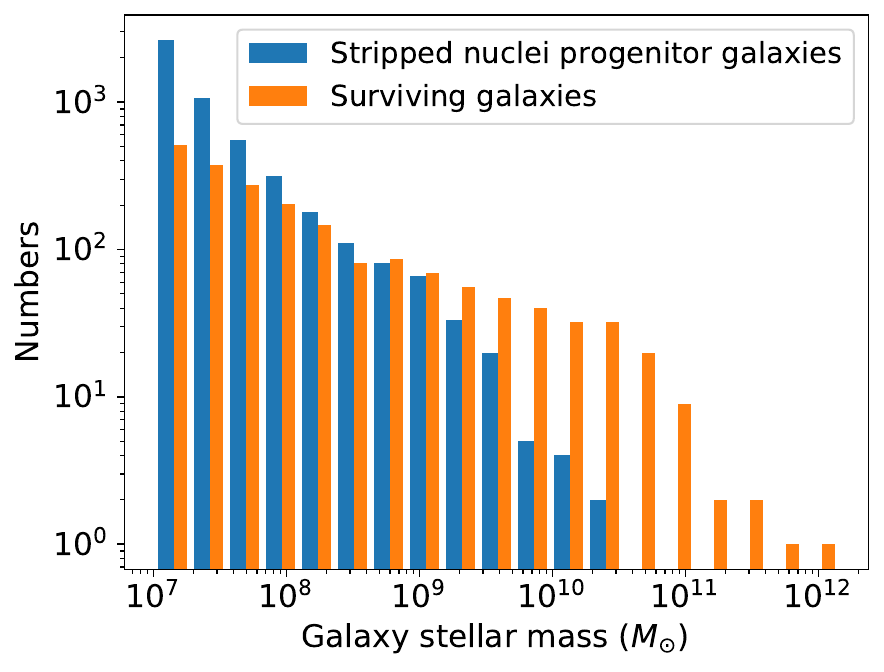}\par 
    \includegraphics[width=\linewidth]{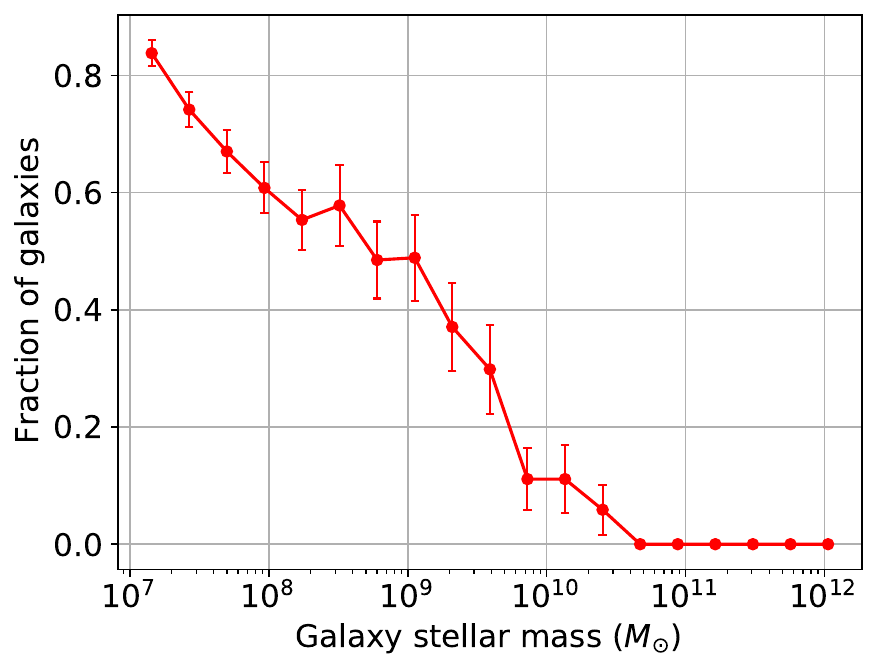}\par 
\end{multicols}
\caption[The number of galaxies at a given stellar mass for stripped nuclei progenitor galaxies and surviving galaxies.]{A comparison of \hl{the stellar masses of stripped nuclei progenitor galaxies} \hlcyan{before merging} with the stellar masses of galaxies that survived to redshift 0. The comparison is made at a given stellar mass for a single simulated cluster. \textbf{Left:} The absolute number of galaxies of each type. \textbf{Right:} The fraction of galaxies in a cluster that were disrupted to become stripped nuclei, defined as $f_{Dis} = N_{Dis}/(N_{Dis}+N_{Sur})$, with uncertainties \hlcyan{propagated as} \mbox{$\sqrt{N}$}}.
\label{fig:nucnums}
\end{figure*}

\begin{table*}
	\centering
	\caption[Ratio of surviving galaxies with black holes to black holes in stripped nuclei]{Ratio of surviving galaxies with black holes to black holes in stripped nuclei for stripped nuclei with black holes M\textsubscript{BH} $>$ 3~$\times$~10\textsuperscript{5}~\(\textup{M}_\odot\) and M\textsubscript{BH} $>$ 10\textsuperscript{6}~\(\textup{M}_\odot\)}
	\label{tab:surviv}
	\begin{tabular}{lccc} 
		\hline
		Cluster-ID & Stripped nuclei/surviving galaxies & Stripped nuclei/surviving galaxies \\
		  & (M\textsubscript{BH} $>$ 10\textsuperscript{6}~\(\textup{M}_\odot\)) & (M\textsubscript{BH} $>$ 3~$\times$~10\textsuperscript{5}~\(\textup{M}_\odot\))\\
		\hline
		28000000000000 & 0.1 & 0.97\\
		28000000000001 & 0.13 & 0.82\\
		28000000000002 & 0.14 & 0.98\\
		28000000000003 & 0.13 & 0.86\\
		28000000000004 & 0.06 & 0.72\\
		28000000000005 & 0.13 & 0.86\\
		28000000000006 & 0.29 & 0.75\\
		\hline
	\end{tabular}
\end{table*}
\hl{The number of galaxies in EAGLE that contain supermassive black holes may depend on the chosen seed mass and seeding mechanism. EAGLE's black hole seeding mechanism involves placing black holes of seed mass \mbox{1.475~$\times$~10\textsuperscript{5}~\(\textup{M}_\odot\)} into dark matter halos with a mass of \mbox{1.475~$\times$~10\textsuperscript{10}~\(\textup{M}_\odot\)}. If these limits were lowered, it would likely increase the number of stripped nuclei found to contain supermassive black holes. Note, however, that the method by which supermassive black holes form and the percentage of observed low-mass galaxies that contain black holes are} \hlcyan{currently} \hl{unclear. This makes it difficult to obtain an accurate fraction of simulated galaxies that would contain black holes at low masses.}
Because of the uncertainty of EAGLE’s black hole masses at lower limits, \hl{partially induced by the seed mass}, we also consider the ratios of progenitor galaxies to surviving galaxies. Figure~\ref{fig:nucnums} plots the number of galaxies of a given stellar mass for surviving galaxies and progenitor galaxies in the EAGLE simulation. We do not factor in nucleation fraction to this plot; however, as we are considering the ratios of galaxies at the same masses, this factor should not have a huge impact.


This plot shows that the ratio of black holes in stripped nuclei to black holes in surviving galaxies depends heavily on the occupation fraction of black holes in low-mass galaxies. Above a stellar mass of 10\textsuperscript{10}~\(\textup{M}_\odot\), where it is believed all galaxies are hosts of supermassive black holes, surviving galaxies outnumber stripped nuclei progenitor galaxies by a factor of 16. In the more unexplored region of 10\textsuperscript{9}~\(\textup{M}_\odot\) to 10\textsuperscript{10}~\(\textup{M}_\odot\), there are more surviving galaxies by a factor of 2. Below a mass of 10\textsuperscript{9}~\(\textup{M}_\odot\), stripped nuclei progenitor galaxies outnumber surviving galaxies. \hl{(Note the limitation of our study to galaxies above a stellar mass of \mbox{10\textsuperscript{7}~\(\textup{M}_\odot\) )}}.

\section{Discussion}

\subsection{Black hole formation models}
\label{section:bhformation}

\hlcyan{Our understanding of how supermassive black holes formed is limited. Different theories involve the direct collapse of protogalactic gas clouds \mbox{\citep{Begelman2006, Latif2013}}, the collapse of supermassive early stars \mbox{\citep{Madau2001, Alvarez2009}}, runaway collisions of stars and/or stellar mass black holes \mbox{\citep{Devecchi2012}}, or even primordial black holes \mbox{\citep{Bean_2002}}. None of these processes can be resolved in large hydrodynamical simulations, therefore, EAGLE uses a seeding mechanism as described in Section \mbox{\ref{sec:eaglebhs}}.}

\hlcyan{\mbox{\citet{Schaye2015}} find that the simulated galaxy stellar mass-black hole mass relation from EAGLE agrees well with observations from \mbox{\citet{McConnell2013}}, although the observations have more scatter. EAGLE also agrees with AGN X-ray luminosity functions \mbox{\citep{RosasGuevara2016}} and observed black hole accretion rates \mbox{\citep{McAlpine2017}}.

In this paper we primarily focus on stripped nuclei with \mbox{M\textsubscript{BH} $>$ 10\textsuperscript{6}~\(\textup{M}_\odot\)}, however, we also work with stripped nuclei that contain a black hole mass more than twice the seed mass
(i.e.  \mbox{M\textsubscript{BH} $>$3~$\times$~10\textsuperscript{5}~\(\textup{M}_\odot\)}). Below a mass of \mbox{M\textsubscript{BH} $>$ 10\textsuperscript{6}~\(\textup{M}_\odot\)}, the black hole content will begin to be represented by only a few particles and therefore becomes unreliable and susceptible to ejection from interactions with other particles due to a lack of dynamical friction \mbox{\citep{Chen2022}}. In addition, \mbox{\citet{Bower_2016}} directly tested the effect of lower black hole seed masses on the EAGLE predictions. They found that while different black hole seed masses converge above \mbox{M\textsubscript{BH} $>$ 10\textsuperscript{6}~\(\textup{M}_\odot\)}, below this mass a lower mass black hole seed mass results in lower mass black holes. Therefore, while we consider the smaller stripped nuclei as hosting black holes, we do not consider the smaller black holes \mbox{(M\textsubscript{BH} $<$ 10\textsuperscript{6}~\(\textup{M}_\odot\))} to have reliable masses.} 

\subsection{The elevated dynamical mass-to-light ratios of UCDs}
\label{section:ucdelevmass}


In Section \ref{section:reselevml}, 
we showed that stripped nuclei containing supermassive black holes have elevated dynamical mass-to-light ratios. In Figure \ref{fig:mieske}, we compare the predicted percentage of stripped nuclei with elevated dynamical mass-to-light ratios to the population of observed UCDs with elevated dynamical masses. We then compared the predicted mean mass elevation ratio of our high mass stripped nuclei to observed UCDs.

\citet{Mieske2013} investigated the possibility that the presence of supermassive black holes causes the elevated dynamical mass-to-light ratios of UCDs. They used a single average measurement of the velocity dispersion to calculate the dynamical masses of UCDs, which is less accurate than models using spatially-resolved spectroscopic data. They calculated the ratio, $\Psi$\textsubscript{obs} = (M/L)\textsubscript{dyn}/ (M/L)\textsubscript{pop} between the dynamical and stellar population M/L ratios, assuming an age of 13 Gyrs for their objects, and a canonical  \citet{Kroupa2001} type mass function for the UCDs. $\Psi$\textsubscript{obs} $> 1$ indicates an elevated dynamical M/L ratio. 
They found that two-thirds of M $>$ 10\textsuperscript{7}~\(\textup{M}_\odot\) UCDs and one-fifth of M $>$ 2~$\times$~10\textsuperscript{6}~\(\textup{M}_\odot\) UCDs have elevated dynamical mass-to-light ratios. The mean mass elevation ratio, $\Psi$\textsubscript{obs} = (M/L)\textsubscript{dyn}/ (M/L)\textsubscript{pop} for observed UCDs with M $>$ 10\textsuperscript{7}~\(\textup{M}_\odot\) was found to be $\Psi$\textsubscript{obs} $= 1.7 \pm 0.2$.

\citet{Voggel_2019} revised the stellar population mass estimates of UCDs, comparing the dynamical masses of UCDs to a new empirical relation between the mass-to-light ratio (M/L) and metallicity. Their results also show that many UCDs have elevated dynamical M/L ratios, with the percentage increasing with mass. They found 0-15 per cent of UCDs with magnitudes fainter than M\textsubscript{V} = -10 mag (M $\approx$2~$\times$~10\textsuperscript{6}~\(\textup{M}_\odot\)), 20-40 per cent of UCDs with magnitudes -10 mag $>$ M\textsubscript{V} $>$ -12 mag (M $\approx$2~$\times$~10\textsuperscript{6}~\(\textup{M}_\odot\) - 10\textsuperscript{7}~\(\textup{M}_\odot\)), and at the brightest magnitudes, 45-80 per cent of UCDs have elevated dynamical M/L ratios. 

In Section \ref{section:lowmass}, we found that in a massive cluster, 100 per cent of stripped nuclei with M\textsubscript{tot} $>$ 10\textsuperscript{7}~\(\textup{M}_\odot\) and $\approx$50 per cent of stripped nuclei with M\textsubscript{tot} $>$ 2~$\times$~10\textsuperscript{6}~\(\textup{M}_\odot\) contain supermassive black holes. Factoring in the percentage of stripped nuclei that contain detectable black holes ($>3$ per cent, Section \ref{section:reselevml}), we find 
$\approx$80-90 per cent of the high mass (M\textsubscript{tot} $>$ 
10\textsuperscript{7}~\(\textup{M}_\odot\)) population contain detectable supermassive black holes that would cause these stripped nuclei to exhibit elevated dynamical mass-to-light ratios. This is consistent with the percentages predicted by \citet{Mieske2013} and \citet{Voggel_2019}. 

We find that $\approx$40-45 per cent of the low mass (M\textsubscript{tot} $>$ 2~$\times$~10\textsuperscript{6}~\(\textup{M}_\odot\)) population of stripped nuclei contain detectable supermassive black holes that would cause these stripped nuclei to exhibit elevated dynamical mass-to-light ratios. This is slightly higher than the predictions from \citet{Mieske2013} and \citet{Voggel_2019}.

Considering that some UCDs likely originated as globular clusters, our results are consistent with all high-mass UCDs being stripped nuclei that harbour supermassive black holes, while a sizeable percentage of low-mass UCDs are likely globular clusters. Supermassive black holes can likely explain all UCDs with elevated dynamical M/L ratios, without other sources such as dark matter or a top or bottom-heavy initial mass function \citep{Dabringhausen2010}.

Approximately 55-60 per cent of M\textsubscript{tot} $>$ 2~$\times$~10\textsuperscript{6}~\(\textup{M}_\odot\) stripped nuclei may be observed as UCDs without elevated dynamical mass-to-light ratios due to the lack of supermassive black holes or with black holes that have masses insufficient to affect their dynamical mass. This result is limited by the lower resolution limit of the EAGLE simulation, where below a mass of 10\textsuperscript{6}~\(\textup{M}_\odot\), black hole masses are unreliable. However, it is worth noting that our results are consistent with many observed UCDs that lack a dynamical indicator of a supermassive black hole still being stripped nuclei.

In Section \ref{section:reselevml}, we calculated the mean mass elevation ratio, $\Psi$\textsubscript{sim} for simulated stripped nuclei. 
We found that the mean $\Psi$\textsubscript{sim} for stripped nuclei with M\textsubscript{tot} $>$ 10\textsuperscript{7}~\(\textup{M}_\odot\) is $\Psi$\textsubscript{sim} $= 1.51^{+0.06}_{-0.04}$. This result is consistent with the results of \citet{Mieske2013}. Figure~\ref{fig:mieske} plots $\Psi$\textsubscript{sim} for our simulated stripped nuclei and $\Psi$\textsubscript{obs} for UCDs from \citet{Mieske2013}. The line of best fit for our data, calculated from Figure~\ref{fig:bhmtotmasslobf} and Figure~\ref{fig:masselev}, is consistent with \citet{Mieske2013}'s data.

\citet{Voggel_2018} noted from the data of \citet{Mieske2013} that there appears to be a trend of lower mass fraction black holes in lower mass UCDs. We also find in Section \ref{section:reselevml} that the mean mass elevation ratio increases with stripped nucleus mass, consistent with their conclusion.

A few factors will affect the comparison of our calculation of mass elevation ratio to the observers' calculations:
\begin{itemize}
\item \citet{Ferr_Mateu_2021} performed a multiwavelength (X-ray, optical spectroscopy, mid-infrared and radio emission) search for black hole activity in UCDs and compact ellipticals. They found evidence for very massive black holes in about 4 per cent of their investigated $\approx$500 UCDs, including examples of UCDs with black holes that potentially exceed our 30 per cent mass fraction limit. Since these results have not been kinematically confirmed, they express caution about using their results as proper values. If this result is confirmed, our 30 per cent upper limit may be too low and should be raised.

\item If supermassive black holes in stripped nuclei experience growth during the merger process that is unaccounted for \hl{by our method of measuring the black hole before merger}, the number of high mass stripped nuclei with M $>$ 10\textsuperscript{6}~\(\textup{M}_\odot\) black holes will be higher. This could affect the mass elevation ratio calculation, although the growth rate is unknown, and the nucleus would also grow, so the impact on the mass elevation ratio is uncertain. 

\item Additionally, there is also the possibility of observations \hl{inaccurately} predicting the mass elevation ratios of UCDs, \hl{since accurate observational measurements of mass elevation ratios require precise calculations of both the dynamical mass and the stellar population mass, which are affected by both observational quality and modelling choices}.
\citet{Voggel_2018} studied the black hole content of UCD 320 and UCD 330 and found that their dynamical M/L ratios had been overestimated in previous studies. 

For UCD 320, \citet{Taylor2010} \hl{used single-slit measurements to} measure $\Psi = 2.5$, while \citet{Mieske2013} \hl{reanalysed the literature data and} found $\Psi = 1.6$. In contrast, \citet{Voggel_2018} found $\Psi = 0.9$, a mass measurement that finds UCD 320 does not have an elevated mass. Their lower $\Psi$ is derived from VLT/SINFONI IFU observations that spatially resolve the UCD, giving a smaller derived effective radius, \hl{and hence a lower dynamical mass} for the UCD than found \hl{by previous studies} and higher stellar population M/L estimates from \hl{a different M/L versus [Fe/H] relation, than that used by the previous studies}.

Similarly, for UCD 330, \citet{Taylor2010} measured $\Psi = 2.3$, while \citet{Mieske2013}'s analysis found $\Psi = 1.7$. In contrast \citet{Voggel_2018} found $\Psi = 0.9$. This lower result stems primarily from \citet{Voggel_2018} VLT/SINFONI IFU observations having a \hlcyan{\mbox{$10 \, \mathrm{km} \, \mathrm{s}^{-1}$}} \hl{lower velocity dispersion (and hence a lower dynamical mass) than previous studies. \mbox{\citet{Taylor2010}}'s measurement was based on reanalyzing data from \mbox{\citet{Rejkuba2007}}, however, their velocity dispersion was \mbox{$\approx$2.5$\sigma$} higher than \mbox{\citet{Rejkuba2007}'}s value. An independent measurement from \mbox{\citet{Hernandez2018}}, is also consistent with \mbox{\citet{Voggel_2018}'s} result, indicating that their velocity dispersion calculation is likely more accurate than that of \mbox{\citet{Taylor2010}}}. 
\hl{The use of a two-component mass model may also have decreased the dynamical mass calculation}. 

\citet{Voggel_2018} found that the \hl{calculations of} \citet{Taylor2010} \hl{result in mass elevation ratios that are too high because of a larger radius for UCD 320 and a higher velocity dispersion for UCD 330}. \hl{Additionally, \mbox{\citet{Ahn2017}} find} a M/L ratio for M59cO that is lower than previous integrated-light measurements \hl{using higher-resolution data}. \hl{These examples} show the difficulty of accurately measuring M/L ratios. There appears, in general, to be a bias towards overestimating M/L ratios measured by integrated-light studies. This could be caused by the \hl{velocity} dispersion \hl{and hence the dynamical mass} being overestimated \hl{via factors such as} galaxy light contaminating the integrated-light spectra of the UCD. \hl{Additionally, errors in the light and mass profile determination of the UCDs, or stellar population modelling choices can also affect the calculated mass elevation ratio}.
If the mass elevation ratios of UCDs in \citet{Mieske2013} are consistently over-estimated, the observed mean mass elevation ratio might be lower. High-quality data is, therefore, key in determining accurate $\Psi$ measurements. 

\hl{Because simulations of the mass elevation ratio rely on simply calculating the dynamical mass of a star cluster with and without a central supermassive black hole present, they are not affected by the challenges of observational modelling.}


\end{itemize}
\hlcyan{In conclusion, a supermassive black hole's presence can elevate a stripped nucleus's dynamical mass.} We find 80-90 per cent of high mass M\textsubscript{tot} $>$ 10\textsuperscript{7}~\(\textup{M}_\odot\) stripped nuclei should have elevated dynamical masses due to central supermassive black holes. This is consistent with the two-thirds of UCDs \hlcyan{observed by} \citet{Mieske2013} and 45-80 per cent of UCDs \hlcyan{observed by} \citet{Voggel_2019} that have elevated dynamical M/L ratios. The overall population of stripped nuclei has a mean mass elevation ratio of $\Psi$\textsubscript{sim} $= 1.51^{+0.06}_{-0.04}$. This is consistent with the value of $\Psi$\textsubscript{obs} $= 1.7 \pm 0.2$ that \citet{Mieske2013} find for observations of UCDs. \hl{These calculations assume that the majority of stripped nuclei with mass \mbox{M\textsubscript{tot} $>$ 10\textsuperscript{7}~\(\textup{M}_\odot\)} host supermassive black holes, however, it is difficult to constrain the percentage that would in reality due to observational limitations.}




\subsection{Massive black holes in low-mass stripped nuclei}
In Section \ref{section:lowmass}, we investigated the possibility of massive black holes \hlcyan{existing in the stripped nuclei population with} total masses overlapping with globular clusters. 
We found that $\approx$0.1 per cent of objects that resemble globular clusters could be stripped nuclei that contain massive black holes. However, our results are consistent with stripped nuclei in the mass range of globular clusters containing no massive black holes. 

Several studies have presented evidence for and against individual globular clusters containing intermediate-mass black holes \citep[e.g.][]{Gebhardt_2002, Ulvestad_2007, Maccarone_2007, Maccarone2008, Noyola_2010, Anderson_2010, Cseh2010, Lutzgendorf_2013, Feldmeier_2013, Lanzoni_2013, Baumgardt_2019}.
At present, there is no solid confirmed evidence for any globular clusters containing supermassive or intermediate-mass black holes.

\citet{Tremou2018} used radio emissions to study 50 Galactic globular clusters and observed no emission consistent with an intermediate-mass black hole and set an upper limit of 10-15 per cent on the fraction of massive globular clusters that could host $\approx$ 10\textsuperscript{4}~\(\textup{M}_\odot\) black holes. While we have no solid black hole mass estimates for the low mass sample of simulated stripped nuclei, even considering the scenario where 100 per cent of stripped nuclei contain supermassive black holes, they would make up no more than a few per cent of the globular cluster population, consistent with observations.

\subsection{Supermassive black holes in galaxy clusters}

One of the strongest motivations for studying the supermassive black hole population in UCDs is that some studies have estimated that the number of supermassive black holes in UCDs could rival the number of supermassive black holes in galaxy nuclei \citep{Seth2014, Voggel_2019}. 

In Section \ref{section:nuclei}, we compared the population of black holes within stripped nuclei to those within surviving galaxy nuclei. In the seven massive galaxy clusters, we found a mean ratio of stripped nuclei black holes/black holes in surviving galaxies of 0.14 for black holes with M\textsubscript{BH} $>$ 10\textsuperscript{6}~\(\textup{M}_\odot\) and 0.85 for black holes with M\textsubscript{BH} $>$ 3~$\times$~10\textsuperscript{5}~\(\textup{M}_\odot\), indicating that black holes in stripped nuclei may not make up a large number of the most massive black holes, but at lower masses, they could rival the number of black holes in galaxy nuclei.

\hl{Note, however, that there are large uncertainties in these calculations due to the uncertain numbers and masses of low-mass black holes in EAGLE, and the fact that the mechanism by which supermassive black holes form is unclear.} Because of the uncertainties in EAGLE's black hole numbers and the fact that we take the stripped nuclei progenitor masses over multiple snapshots, we also considered the ratios of surviving galaxies to disrupted galaxies. \hl{Low-mass black hole fractions in galaxies are currently poorly constrained due to limited observations of black holes in low-mass galaxies.}


Below a stellar mass of 10\textsuperscript{9}~\(\textup{M}_\odot\), we find three times as many stripped nuclei progenitor galaxies as surviving galaxy nuclei. The occupation fraction of black holes in these low mass galaxies is unknown, but $\approx$100 per cent is possible \citep{Miller2015}. If this is the case, the number of intermediate or supermassive black holes in stripped nuclei could outnumber those in galaxy nuclei. 

If we assume 100 per cent of nucleated galaxies with M\textsubscript{*} $>$ 10\textsuperscript{7}~\(\textup{M}_\odot\) host intermediate or supermassive black holes, and factor in the fraction of galaxies in each mass range \hlcyan{that} we expect host nuclei, we find the number of black holes in stripped nuclei outnumbers those in galaxy nuclei by a ratio of 1.2. The masses of black holes within these galaxies are likely $>$ 10\textsuperscript{3}~\(\textup{M}_\odot\) \citep[e.g.][]{Barai_2018,Mezcua_2018}. 

However, even if we assume lower occupation fractions, a sizeable number of black holes in galaxy clusters could reside in stripped nuclei. \citet{Greene_2020} review the current search for intermediate-mass black holes and conclude that in the stellar mass range of 10\textsuperscript{9} to 10\textsuperscript{10}~\(\textup{M}_\odot\), over 50 per cent of galaxies should contain black holes. Using this lower limit, along with the assumption that M\textsubscript{*} $>$ 10\textsuperscript{10}~\(\textup{M}_\odot\) galaxies have a 100 per cent occupation fraction, and M\textsubscript{*} $<$ 10\textsuperscript{9}~\(\textup{M}_\odot\) galaxies have a 0 per cent black hole occupation fraction, we find 53 stripped nuclei should contain supermassive black holes (with masses likely above 10\textsuperscript{5}~\(\textup{M}_\odot\)) and 187 supermassive black holes in surviving galaxies. Therefore, considering the minimum known constraints on supermassive black hole occupation fractions, stripped nuclei should represent a minimum increase in the supermassive black hole numbers in galaxy clusters of 30 per cent. 

Similar results are found with predictions from \hl{cosmological} simulations and semi-analytic models. \citet{Pacucci_2021} summarized the predicted occupation fractions of black holes in dwarf galaxies from various studies in their Figure 3. \hlcyan{Based on these studies,} we calculate the impact stripped nuclei would have on the relative number of black holes in galaxy clusters.

\citet{Ricarte_2017} used a semi-analytic model to predict the growth of black holes. Their optimistic occupation fraction would result in a 57 per cent increase in black holes in galaxy clusters from stripped nuclei, and their pessimistic prediction 28 per cent.
\citet{Bellovary_2018} studied the black holes in dwarf galaxies in high-resolution zoom-in simulations, \hl{run with the state-of-the-art N-body
Tree+SPH code ChaNGa. They modelled black hole formation via the direct collapse scenario, where black holes with seed mass \mbox{2.5~$\times$~10\textsuperscript{5}~\(\textup{M}_\odot\)} form in dense \mbox{($n_H >$ 1.5~$\times$~10\textsuperscript{4} cm\textsuperscript{-3}}) areas
for the higher resolution runs and black holes with \mbox{5~$\times$~10\textsuperscript{5}~\(\textup{M}_\odot\)} seed mass form in dense \mbox{($n_H >$ 3000 cm\textsuperscript{-3})} areas for the 
lower-resolution runs. }Using their prediction, we would find a 97 per cent increase in black holes in galaxy clusters, \hl{similar to the seed mass predictions from EAGLE}. 

\citet{Askar_2022} modelled IMBH growth via the merger of stellar clusters, factoring in the likelihood of the black hole being kicked from the cluster during a gravitational wave merger. In the high kick scenario, their occupation fraction calculation would result in a 57 per cent increase in black holes in galaxy clusters, and the low kick scenario results in an 89 per cent increase. 

So, in summary, semi-analytic and simulation predictions give a 28-97 per cent increase in black holes in galaxy clusters from stripped nuclei, consistent with the predictions from UCD observations. 


A complicating factor in determining the relative numbers of supermassive black holes in surviving galaxies as compared to disrupted galaxies is that the ratio between black hole mass and galaxy stellar mass may not stay constant between the time at which the stripped nuclei progenitor galaxies are disrupted and the present day. 

Several studies have investigated how the ratio of black hole to galaxy stellar mass evolves with redshift and found a higher black hole to galaxy mass ratio at redshift 1-4 \citep[e.g.][]{Merloni2010, Willott_2013, Decarli_2018} 
up to a possible ten times discrepancy at z $>$ 6.
However, other studies, \citep[e.g.][]{Jahnke_2009,Cisternas_2011} found no evidence for an
evolution in the black hole to host galaxy mass ratio, and there may be selection bias involved because high-redshift studies focus on exceptionally bright galaxies that host more massive black holes \citep[e.g.][]{Decarli_2018}.

While galaxy mergers that form stripped nuclei continue to redshift zero, they primarily merge in the universe's early stages. In Paper II we found that the mean merger time is 9.0 $\pm$ 0.2 Gyr or z $\approx$ 1.3. Many progenitor galaxies will have formed and merged at even higher redshifts. These galaxies may have overmassive black holes relative to their stellar masses, resulting in the stripped nuclei progenitor galaxies containing higher mass black holes relative to the surviving galaxies and having a higher black hole occupation fraction. Figure~\ref{fig:obssim} shows that the black hole masses of stripped nuclei progenitors with M\textsubscript{BH} $>$ 10\textsuperscript{6} \(\textup{M}_\odot\) sit slightly above the black hole masses in surviving simulated galaxies, supporting the fact that black holes in older galaxies are overmassive relative to those at z = 0.

\citet{Seth2014} compared the UCD population to the population of galaxies in the Fornax cluster that are likely to host supermassive black holes. Considering just high mass UCDs, they concluded that they would represent a $\approx$40 per cent increase over the number of galaxy black holes in Fornax. Including low-mass UCDs with elevated dynamical M/L ratios, UCDs may more than double the number of black holes in the Fornax cluster. Their result is consistent with our predictions.

\citet{Voggel_2019} estimated the occupation probability of SMBHs in UCDs as a function of their luminosity based on modelling the dynamical M/L ratios of UCDs. They then compared these UCDs to galaxy nuclei in the Fornax and Virgo clusters. They found a ratio of SMBH UCDs to current nuclei of $0.89^{+0.57}_{-0.32}$ in Fornax and $0.83^{+0.22}_{-0.37}$ in Virgo, indicating that supermassive black holes in UCDs may equal or outnumber those in galaxy nuclei. Their result is consistent with our finding that the mean ratio of black holes in stripped nuclei to those in surviving galaxies is 0.85 for black holes with M\textsubscript{BH} $>$ 3~$\times$~10\textsuperscript{5}~\(\textup{M}_\odot\).

\hl{Overall, there remain large uncertainties on the numbers and masses of black holes in low-mass galaxies, which constrain predictions of the numbers and masses of black holes in stripped nuclei from both simulations and observations. In particular, there are large uncertainties on the predicted number of black holes in stripped nuclei}. We conclude that even in a pessimistic scenario, assuming a lower bound on the number of galaxies expected to contain supermassive black holes and no evolution of the stellar mass-black hole mass relation, stripped nuclei should represent a 30 per cent increase in the number of black holes in galaxy clusters. Assuming higher black hole occupation fractions, or considering the black hole numbers predicted by EAGLE, stripped nuclei could equal or outnumber black holes in galaxy nuclei, consistent with predictions from \citet{Seth2014} and \citet{Voggel_2019}.

On average, the masses of the black holes in stripped nuclei will be lower than those in galaxy nuclei due to being predominately from lower-mass galaxies. However, increased black holes in galaxy clusters would increase expected tidal disruption events and binary black hole merger rates.

\section{Summary}

This paper presents the first work to predict the population of supermassive black holes in stripped nuclei using a \hlcyan{cosmological} hydrodynamic simulation. We find that the majority of stripped nuclei may contain supermassive black holes, the black holes in stripped nuclei can explain those in observed UCDs, our predicted mass elevation ratio of simulated stripped nuclei is consistent with that of observed UCDs, and supermassive black holes in stripped nuclei could equal or outnumber those in normal galaxy nuclei. Our results are summarised as follows:
\begin{enumerate}
  \item The simulation can reproduce the expected black hole masses of observed UCDs, a test of our model of UCD formation establishing that the simulation can accurately predict the population of black holes in stripped nuclei.
  \item Approximately 27 per cent of stripped nuclei with \hl{total mass (stellar mass plus black hole mass, see Table 1)}, M\textsubscript{tot} $>$ 10\textsuperscript{7}~\(\textup{M}_\odot\) host  M\textsubscript{BH} $>$ 10\textsuperscript{6}~\(\textup{M}_\odot\) supermassive black holes. Approximately 51 per cent of stripped nuclei with M\textsubscript{tot} $>$ 2~$\times$~10\textsuperscript{6}~\(\textup{M}_\odot\) host  M\textsubscript{BH} $>$ 3~$\times$~10\textsuperscript{5}~\(\textup{M}_\odot\) supermassive black holes.
  \item Approximately 2-4 per cent of stripped nuclei in the mass range of globular clusters could host intermediate or supermassive black holes, but our results are consistent with no globular cluster mass stripped nuclei hosting black holes.
  \item Approximately 80-90 per cent of supermassive black holes within stripped nuclei make up more than $\approx$3 per cent of the stripped nuclei mass and hence should be dynamically detectable under good observing conditions. This is consistent with the percentage of high-mass UCDs that are observed to have elevated dynamical masses.
  \item Our predicted mass elevation ratio of simulated stripped nuclei with M\textsubscript{tot} $>$ 10\textsuperscript{7}~\(\textup{M}_\odot\) is $\Psi$\textsubscript{sim} $= 1.51^{+0.06}_{-0.04}$. This is consistent with observed UCDs, which have $\Psi$\textsubscript{obs} = $1.7 \pm 0.2$ \citep{Mieske2013}. \hl{This calculation assumes that the majority of these high-mass stripped nuclei contain black holes.}
  \item The mean ratio of black holes in stripped nuclei to black holes in surviving galaxies is 0.14 for black holes with M $>$ 10\textsuperscript{6}~\(\textup{M}_\odot\) and 0.85 for black holes with M $>$ 3~$\times$~10\textsuperscript{5}~\(\textup{M}_\odot\). \hl{Note that this prediction depends on the unreliable numbers of low-mass black holes in the EAGLE simulation.}
  \item At high stellar masses, we find surviving galaxies outnumber progenitor galaxies of stripped nuclei. At low stellar masses, progenitor galaxies outnumber surviving galaxies. Therefore the ratio of black holes in stripped nuclei to surviving galaxies is dependent on the occupation fraction of black holes in low-mass galaxies.
  \item If we assume pessimistic occupation fractions of supermassive black holes in galaxies, stripped nuclei should represent a minimum increase in supermassive black holes in galaxy clusters of 30 per cent. Higher occupation fractions could result in the number of supermassive black holes in stripped nuclei equalling or outnumbering those in galaxies.
\end{enumerate}

\section*{Acknowledgements}
This work used the DiRAC@Durham facility managed by the Institute for Computational Cosmology on behalf of the STFC DiRAC HPC Facility (www.dirac.ac.uk). The equipment was funded by BEIS capital funding via STFC capital grants ST/K00042X/1, ST/P002293/1, ST/R002371/1 and ST/S002502/1, Durham University and STFC operations grant ST/R000832/1. DiRAC is part of the National e-Infrastructure.

JP was supported by the Australian government through the Australian Research Council's Discovery Projects funding scheme (DP200102574).
\section*{DATA AVAILABILITY STATEMENT}

Data available on request.





\bibliographystyle{mnras}
\bibliography{biblio} 








\bsp	
\label{lastpage}
\end{document}